\documentclass[authoryear,review,preprint,12pt]{cas-sc}

\usepackage[authoryear]{natbib}

\usepackage{subcaption}
\usepackage{hyperref}

\usepackage{aas_macros}
\newcommand*\degr{\ensuremath{^\circ}}
\begin{document}
\let\WriteBookmarks\relax
\def\floatpagepagefraction{1}
\def\textpagefraction{.001}
\shorttitle{A Global Fireball Observatory}
\shortauthors{Devillepoix et~al.}

\title [mode = title]{A Global Fireball Observatory}

\author[1]{H. A. R. Devillepoix} 
\ead{hadrien.devillepoix@curtin.edu.au}

\author[1]{M. Cup{\'a}k} 
\author[1]{P. A. Bland} 
\author[1]{E. K. Sansom} 
\author[1]{M. C. Towner} 
\author[1]{R. M. Howie} 
\author[1]{B. A. D. Hartig} 
\author[1]{T. Jansen-Sturgeon} 
\author[1]{P. M. Shober} 
\author[1]{S. L. Anderson} 
\author[1]{G. K. Benedix} 
\author[1]{D. Busan} 
\author[1]{R. Sayers} 

\address[1]{Space Science and Technology Centre, School of Earth and Planetary Sciences, Curtin University, GPO Box U1987, Perth WA 6845, Australia}

\author[2,3]{P. Jenniskens} 
\address[3]{NASA Ames Research Center, Moffett Field, California 94035, USA}
\address[2]{SETI Institute, Carl Sagan Center, Mountain View, California 94043, USA}

\author[2]{J. Albers}

\author[4]{C. D. K. Herd} 
\author[4]{P. J. A. Hill} 
\address[4]{Dept. of Earth and Atmospheric Sciences, 1-26 Earth Sciences Building, University of Alberta, Edmonton, Alberta, T6G 2EG, Canada}

\author[5,6]{P. G. Brown} 
\author[5]{Z. Krzeminski}
\address[5]{Dept. of Physics and Astronomy, University of Western Ontario, London, ON, Canada N6 A 3K7}
\address[6]{Dept. Earth Sciences, University of Western Ontario, London, ON, Canada N6 A 3K7}

\author[7]{G. R. Osinski} 
\address[7]{Institute for Earth and Space Exploration, University of Western Ontario, London, ON, Canada N6 A 3K7}

\author[8]{H. {Chennaoui Aoudjehane}} 
\address[8]{GAIA Laboratory, Hassan II University of Casablanca, Faculty of Sciences Ain Chock, km 8 Route d’El Jadida, 20150 Casablanca, Morocco}

\author[9]{Z. Benkhaldoun} 
\author[9]{A. Jabiri}
\author[9]{M. Guennoun}
\author[9]{A. Barka}
\address[9]{Ouka\"{\i}meden Observatory, LPHEA, Cadi Ayyad University, Marrakech, Morocco}

\author[10]{H. Darhmaoui} 
\address[10]{School of Science and Engineering, Al Akhawayn University in Ifrane, Morocco 53000}

\author[11, 1, 12]{L. Daly} 
\address[11]{School of Geographical and Earth Sciences, University of Glasgow, Glasgow, G12 8QQ, United Kingdom}
\address[12]{Australian Centre for Microscopy and Microanalysis, University of Sydney, Sydney 2006, NSW, Australia}

\author[13]{G. S. Collins} 
\author[13]{S. McMullan} 
\address[13]{Dept. Earth Science and Engineering, Imperial College, London, SW7 2AZ, United Kingdom}
\author[14]{M. D. Suttle} 
\address[14]{Dipartimento di Scienze della Terra, Universita di Pisa, 56126 Pisa, Italy}

\author[15]{T. Ireland} 
\address[15]{Planetary Science Institute, The Australian National University, Canberra, ACT 2611, Australia}
\author[16]{G. Bonning} 
\author[16]{L. Baeza} 
\address[16]{Research School of Earth Sciences, The Australian National University, Canberra, ACT 2611, Australia}

\author[17]{T. Y. Alrefay} 
\address[17]{National Center for Astronomy, KACST, Riyadh, KSA}

\author[18]{J. Horner} 
\address[18]{Centre for Astrophysics, University of Southern Queensland, Toowoomba, Queensland 4350, Australia}

\author[19]{T. D. Swindle} 
\author[19]{C. W. Hergenrother} 
\address[19]{Lunar and Planetary Laboratory, University of Arizona, 1629 E University Boulevard Tucson, Arizona 85721, USA}

\author[20]{M. D. Fries} 
\address[20]{Astromaterials Research and Exploration Science, NASA Johnson Space Center, Houston, Texas 77058, USA}

\author[21]{A. Tomkins} 
\author[21]{A. Langendam} 
\address[21]{School of Earth, Atmosphere and Environment, Melbourne, Victoria, Australia}

\author[22]{T. Rushmer} 
\author[22]{C. O'Neill} %
\address[22]{Department of Earth and Planetary Sciences, Macquarie University, North Ryde, Sydney, NSW 2109, Australia}

\author[23]{D. Janches} 
\address[23]{ITM Physics Laboratory, Heliophysics Science Division, GSFC/NASA, Greenbelt, MD 20771, USA}
\author[24]{J. L. Hormaechea} 
\address[24]{Facultad de Ciencias Astronómicas y Geof\'{\i}sicas, Universidad Nacional de La Plata, Argentina ; Estacion Astronomica Rio Grande, Rio Grande, Tierra del Fuego, Argentina}

\author[25]{C. Shaw} 
\address[25]{Mullard Radio Astronomy Observatory, University of Cambridge, Lord's Bridge, Barton, Cambridge CB3 7EX, United Kingdom}

\author[26]{J. S. Young} 
\address[26]{Cavendish Laboratory, University of Cambridge, Cambridge CB3 0HE, United Kingdom}

\author[27]{M. Alexander}
\address[27]{Galloway Astronomy Centre, Glasserton, Nr Whithorn, Scotland, DG8 8NE, United Kingdom}

\author[28]{A. D. Mardon}
\address[28]{Newby Hall and Gardens, Ripon, Yorkshire, HG4 5AE, United Kingdom}

\author[29]{J. R. Tate}
\address[29]{The Spaceguard Centre, Llanshay Lane, Knighton, Powys, LD7 1LW, United Kingdom.}

\begin{abstract}
The world's meteorite collections contain a very rich picture of what the early Solar System would have been made of, however the lack of spatial context with respect to their parent population for these samples is an issue.
The asteroid population is equally as rich in surface mineralogies, and mapping these two populations (meteorites and asteroids) together is a major challenge for planetary science.
Directly probing asteroids achieves this at a high cost.
Observing meteorite falls and calculating their pre-atmospheric orbit on the other hand, is a cheaper way to approach the problem.
The Global Fireball Observatory (GFO) collaboration was established in 2017 and brings together multiple institutions (from Australia, USA, Canada, Morocco, Saudi Arabia, the UK, and Argentina) to maximise the area for fireball observation time and therefore meteorite recoveries.
The members have a choice to operate independently, but they can also choose to work in a fully collaborative manner with other GFO partners.
This efficient approach leverages the experience gained from the Desert Fireball Network (DFN) pathfinder project in Australia.
The state-of-the art technology (DFN camera systems and data reduction) and experience of the support teams is shared between all partners, freeing up time for science investigations and meteorite searching.
With all networks combined together, the GFO collaboration already covers 0.6\% of the Earth's surface for meteorite recovery as of mid-2019, and aims to reach 2\% in the early 2020s.
We estimate that after 5 years of operation, the GFO will have observed a fireball from virtually every meteorite type.
This combined effort will bring new, fresh, extra-terrestrial material to the labs, yielding new insights about the formation of the Solar System.
\end{abstract}

\begin{keywords}
meteoroids \sep meteors \sep asteroids: general
\end{keywords}

\maketitle

\newpage

\section{Introduction}
Our view of the early solar system is provided by the variety of meteorites that fall to Earth each year.
These meteorites tell of diverse processes affecting the building of planetesimals and planets in the earliest stages of our solar system.
Our interpretation of the messages provided is clouded by a lack of constraint in where and when these processes are occurring. 
Meteorites are sourced primarily from bodies which are in Earth crossing orbits and from the Main Asteroid Belt.
Spectral analysis of asteroids shows diverse surface compositions that can be related to the mineralogies of meteorites.
Still, relating meteorites to potential source bodies is difficult because of space weathering of asteroids, and the close relationships of many meteorites.
Direct mapping of meteorites to a particular asteroid is being achieved through remote analysis and sample-return missions, which achieve the goals at extremely high cost (e.g. JAXA's \textit{Hayabusa-1} and \textit{Hayabusa-2}, NASA's \textit{OSIRIS-REx} and \textit{Stardust} missions).
A more complete overview of the relationships between meteorites and asteroids is a major challenge for planetary science \citep{2015aste.book...43R}.

A cheaper approximation is to observe meteorite falls with enough accuracy to calculate their pre-atmospheric orbit and the location of the meteorites on the ground for recovery.
This has been done for over 50 years, but at a low success rate given the frequency of meteorite dropping events within the deployed camera coverage \citep{1989Metic..24...65H,1998M&PS...33...49O}.
Although the number of successful meteorite recoveries has increased significantly in the last 10-15 years \citep{2015aste.book..257B}, there is still a significant deficit of meteorites found compared to the number of meteorite producing fireballs that are observed.
Systematic issues in the way observation data are analysed may contribute to this discrepancy \citep{2014A&A...570A..39S}, but locating meteorites outside of populated areas is a non-trivial task and is likely the main limiting factor.  
The recovery rate could be increased by improving search techniques, which generally involve small teams conducting visual searches on foot (for the most part).
Another solution to increase the global number of meteorites recovered is making the collecting area larger, in order to observe more falls \citep{2017ExA...tmp...19H}.

Conversely, in some parts of the world visually observed meteorites falls are routinely recovered without the help of specific fireball observing equipment.
Over the last 20 years, a number of meteorites falls have been recovered in Morocco thanks to the considerable local interest and awareness in meteorites, relying on intuitive searching methodology \citep{2012Sci...338..785A,2019LPICo2157.6297C},
effectively making this part of the world the area where the largest number of falls are recovered per unit of surface area \citep{2016LPICo1921.6119C}.
In the USA, a number of falls have been recovered in recent years without the aid of detailed fireball observations, using Doppler radar signatures of the falling meteorites (more on this in Sec. \ref{sec:Doppler}).
For these specific areas, deploying fireball observation hardware is merely an easy way of adding value (orbital context) to the samples that are already being recovered.

From the 30-40 meteorite cases for which an orbit has been derived, some clues about where the most common types of meteorites come from are already starting to emerge.
Orbits of LL chondrites seem to point to a source in the inner edge of the inner main belt.
Combined with Cosmic-Ray Exposure (CRE) ages, the dynamical history of recovered H and L chondrites indicate that there might be multiple sources for these groups.
CM chondrites likely come from a source that can efficiently feed material into the 3:1 mean-motion resonance with Jupiter.
The reader is referred to \citet{2014me13.conf...57J,2020IAUGA..30....9J} and references therein for a more in-depth review.

In this benchmark paper we describe how a global collaborative approach to fireball observation and meteorite recovery will help build a geological map of the inner Solar System as well as a better understanding of the flux of centimetre to metre scale impactors on Earth.

\section{The Global Fireball Observatory collaboration}\label{sec:gfo_collab}

The Global Fireball Observatory (GFO) collaboration was established in 2017 thanks to support from the Australian Research Council Linkage Infrastructure, Equipment and Facilities (LIEF) program.
The goal is to deploy fireball observatories all around the world to maximise fireball observation area and therefore meteorite recoveries, using common hardware and data reduction, and facilitate collaboration amongst a range of planetary science partner institutions.
The project brings together over 19 institutions within Australia, USA, Canada, Morocco, Saudi Arabia, UK, and Argentina (Tab. \ref{table:subnets}). A key aspect of the project is that partners have independence.
Each partner has a discrete network which is operated entirely independently as a distinct regional or national network.
The GFO itself can be thought of as an emergent "network-of-networks", akin to a large-scale astronomy facility, allowing datasets from partner networks to be consolidated and analysed with common processing.
The project builds upon the engineering heritage from the Desert Fireball Network (DFN) pathfinder project in Australia in order to roll out operational partner networks as quickly as possible.

A few systems were installed in California prior to the full scale global deployment and establishment of the GFO collaboration.
This effort soon paid off with the first GFO success---the recovery of the Creston meteorite in California in 2015 \citep{2019M&PS...54..699J}. The one-design approach for the camera hardware (see Sec. \ref{sec:hardware}) brings economies of scale on hardware design and building costs, and makes it easier to develop and maintain an automated data reduction pipeline.
This approach also allows research groups that are not necessarily involved with fireball studies to broaden their range of expertise quickly and at a relatively low cost.

\begin{figure*}
		\begin{subfigure}{.75\textwidth}
			\centering
			\includegraphics[width=\linewidth]{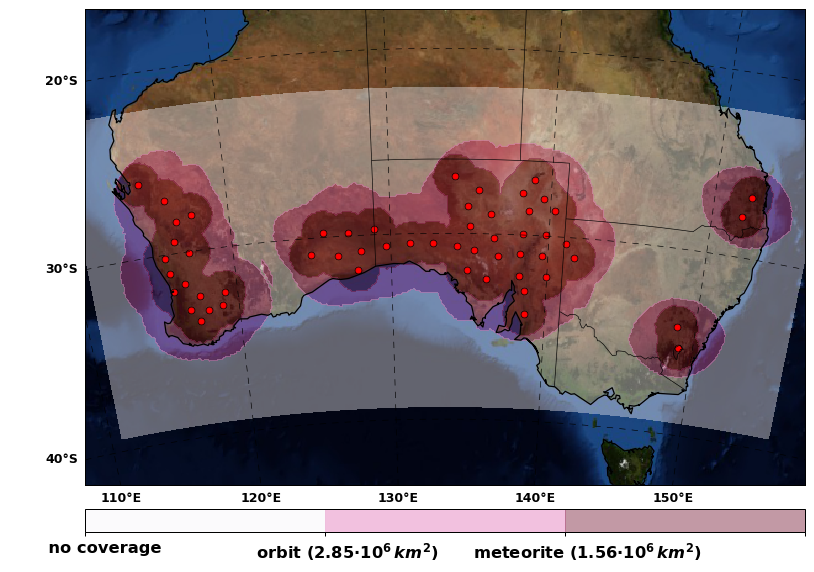}
		\end{subfigure}%
		\begin{subfigure}{.25\textwidth}
			\centering
			\includegraphics[width=\linewidth]{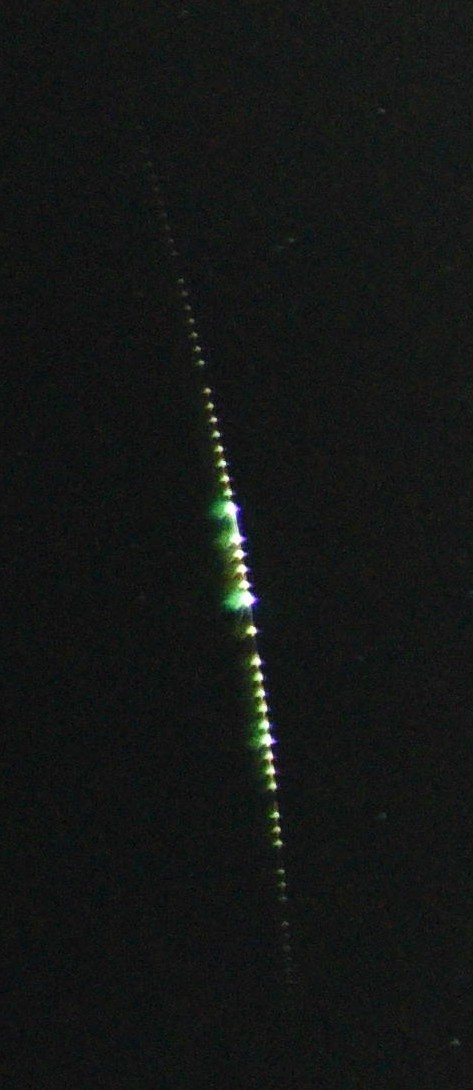}
			\label{fig:aus_fireball_pic}
		\end{subfigure}
		\caption{GFO networks in Australia: Desert Fireball Network (Western and South Australia), South-Eastern Australian Fireball Network, and Queensland Fireball Network.
		left: Fireball observation coverage in Australia as of January 2020. Each red dot corresponds to an observatory site. See Sec. \ref{sec:gfo_collab} for explanation on coverage area.
		right: Fireball observed from Mount Stromlo observatory near Canberra on Sept 9, 2019.}
		\label{fig:aus_fig}
	\end{figure*}

\begin{figure*}
		\begin{subfigure}{.55\textwidth}
			\centering
			\includegraphics[width=\linewidth]{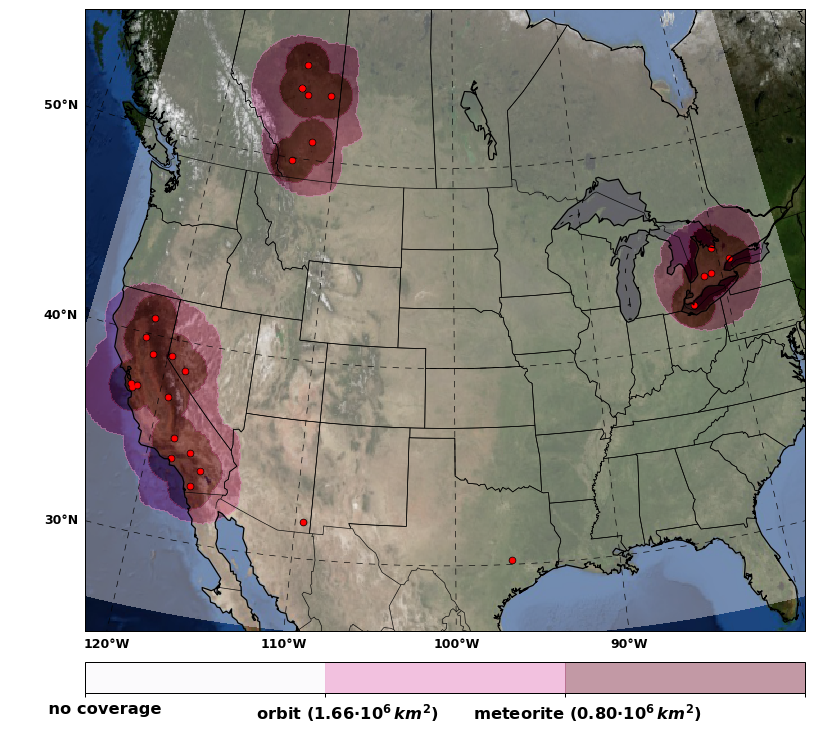}
		\end{subfigure}%
		\begin{subfigure}{.45\textwidth}
			\centering
			\includegraphics[width=\linewidth]{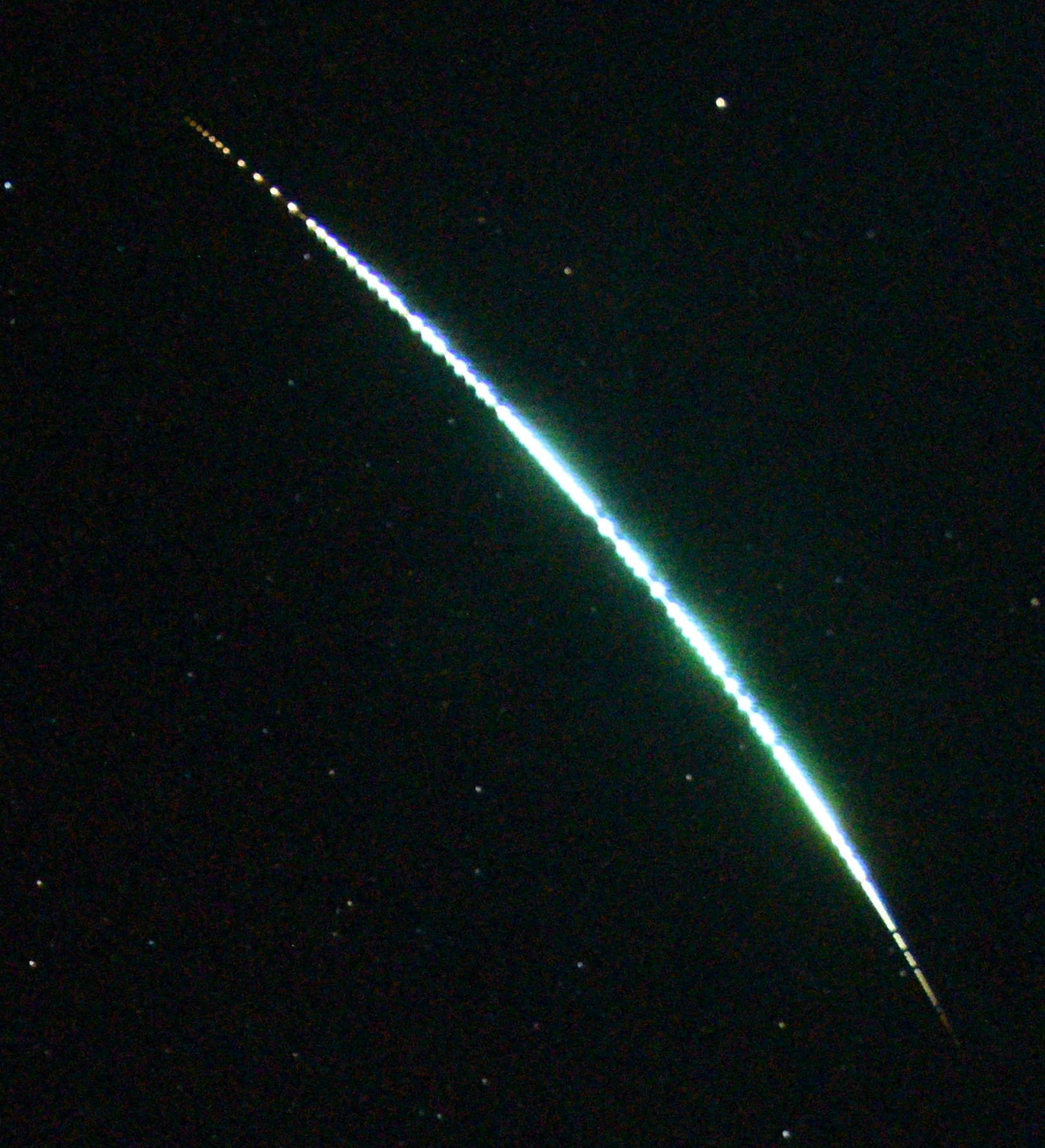}
			\label{fig:northamerica_fireball_pic}
		\end{subfigure}
		\caption{The GFO networks in North America: NASA Meteorite Tracking and Recovery Network (California and Nevada), Meteorite Observation and Recovery Project 2.0 (Alberta), Southern Ontario Meteor Network (SOMN), Arizona Fireball Network.
		left: Fireball observation coverage in North America as of January 2020. Each red dot corresponds to an observatory site. See Sec. \ref{sec:gfo_collab} for explanation on coverage area.
		right: Fireball observed from the Allen Telescope Array station in California on Jul 1, 2019.}
		\label{fig:northamerica_fig}
	\end{figure*}

\begin{figure*}
		\begin{subfigure}{.58\textwidth}
			\centering
			\includegraphics[width=\linewidth]{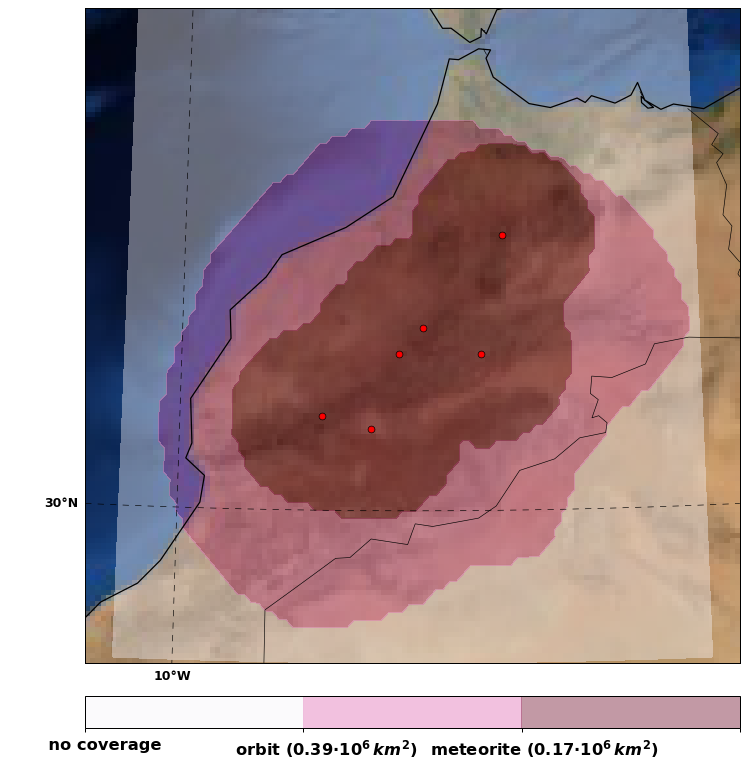}
		\end{subfigure}%
		\begin{subfigure}{.42\textwidth}
			\centering
			\includegraphics[width=\linewidth]{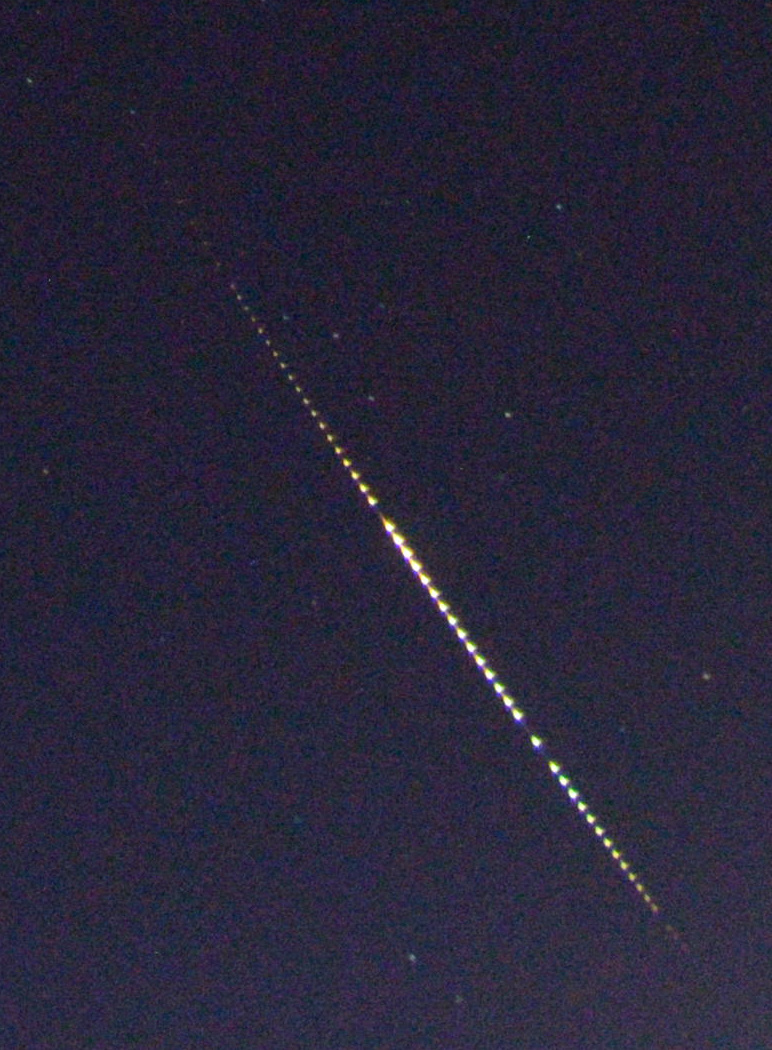}
			\label{fig:morocco_fireball_pic}
		\end{subfigure}
		\caption{The Moroccan Observatory for Fireball Detections (MOFID).
		left: Fireball observation coverage in Morocco as of January 2020. Each red dot corresponds to an observatory site. See Sec. \ref{sec:gfo_collab} for explanation on coverage area.
		right: Fireball observed from Ouka\"{\i}meden Observatory in Morocco on Dec 25, 2018.}
		\label{fig:morocco_fig}
	\end{figure*}

\begin{figure*}
		\begin{subfigure}{.55\textwidth}
			\centering
			\includegraphics[width=\linewidth]{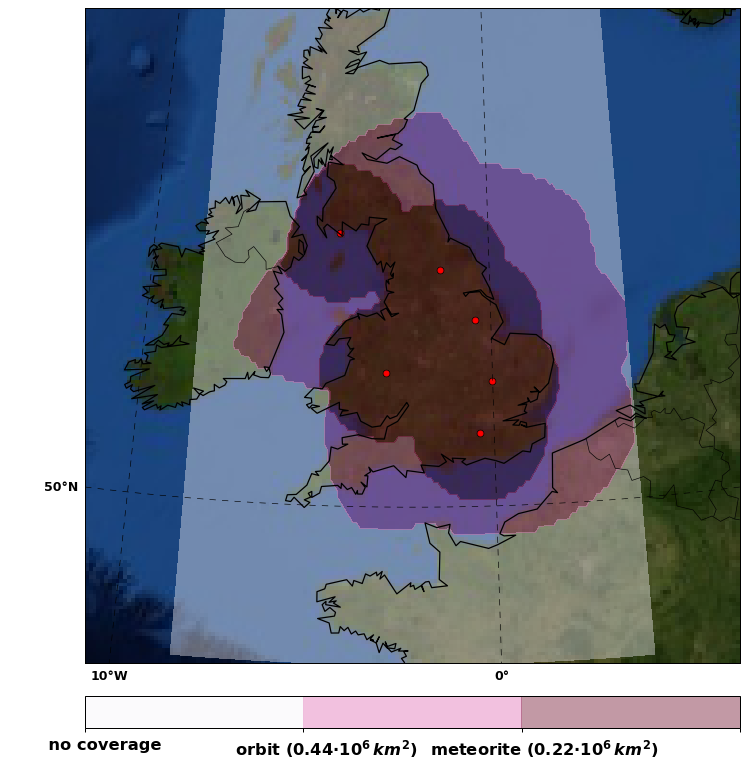}
		\end{subfigure}%
		\begin{subfigure}{.45\textwidth}
			\centering
			\includegraphics[width=\linewidth]{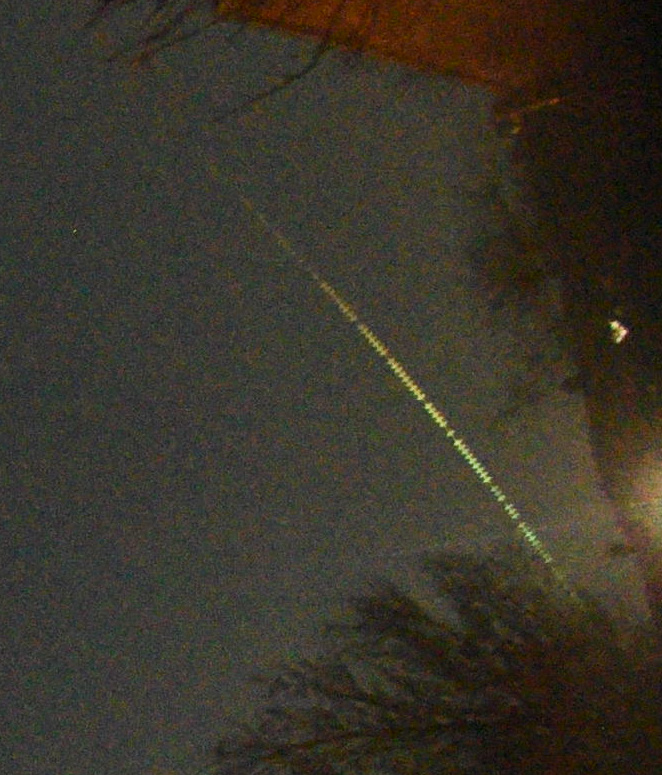}
			\label{fig:uk_fireball_pic}
		\end{subfigure}
		\caption{The UK Fireball Network (UKFN).
		left: Fireball observation coverage in the UK as of January 2020. Each red dot corresponds to an observatory site. See Sec. \ref{sec:gfo_collab} for explanation on coverage area.
		right: Fireball observed from Welwyn station in the South-East of the UK on Feb 15, 2019.}
		\label{fig:fig_UK}
	\end{figure*}

\begin{figure*}
		\begin{subfigure}{.55\textwidth}
			\centering
			\includegraphics[width=\linewidth]{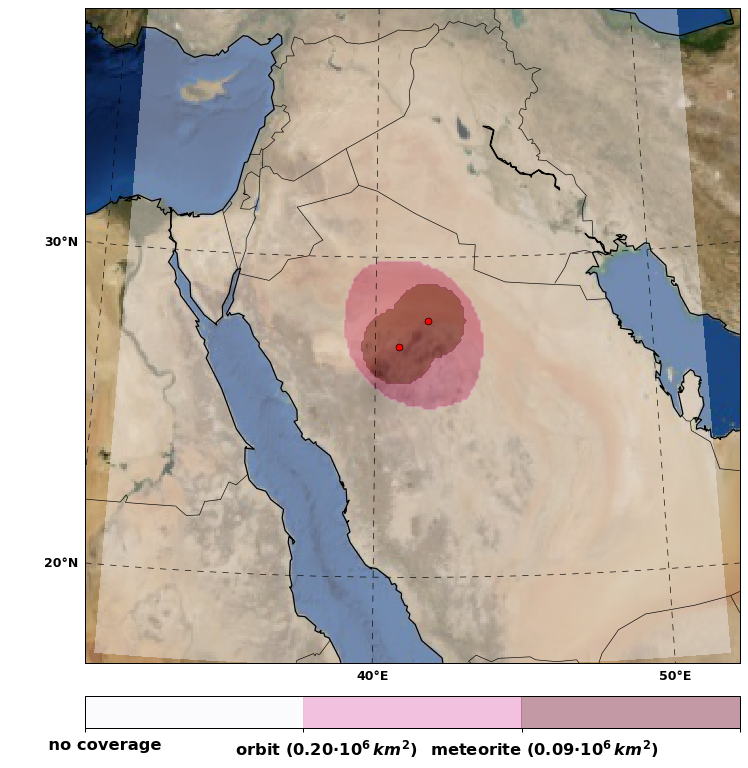}
		\end{subfigure}%
		\begin{subfigure}{.45\textwidth}
			\centering
			\includegraphics[width=\linewidth]{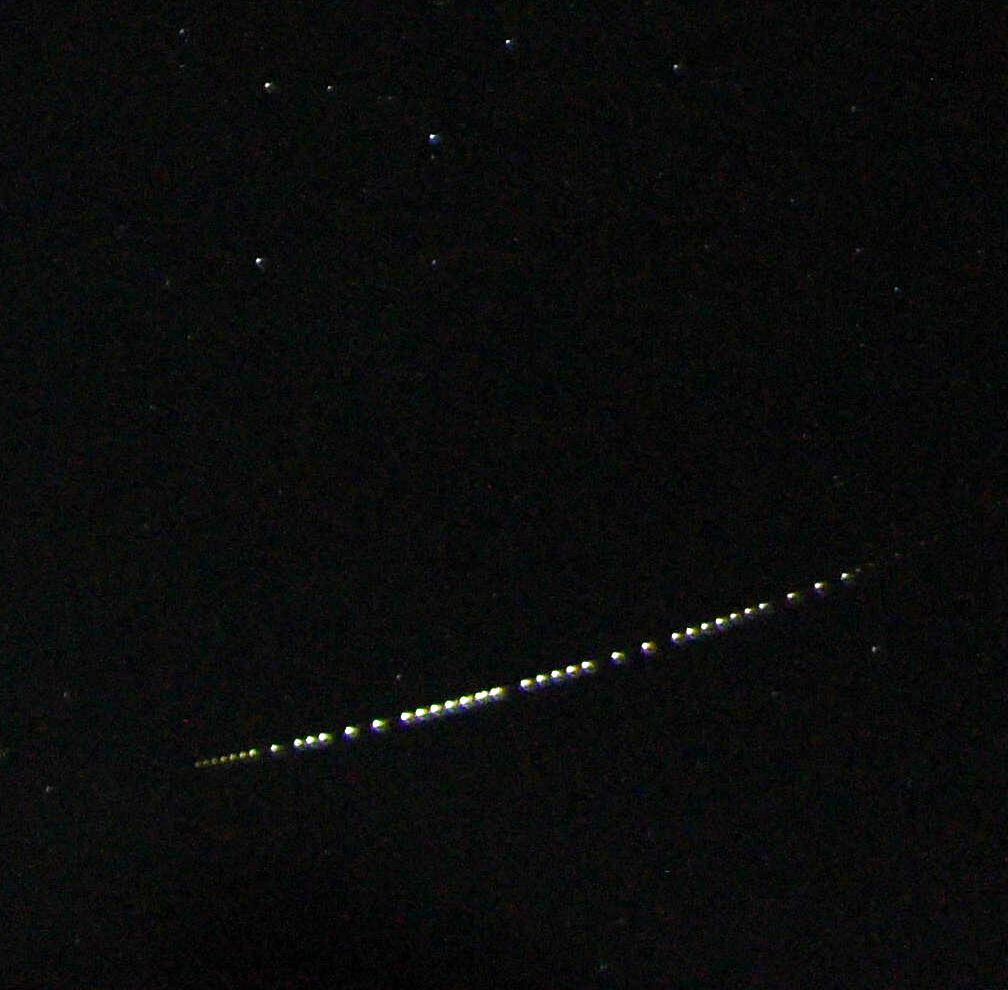}
			\label{fig:Middle_East_fireball_pic}
		\end{subfigure}
		\caption{The Kingdom of Saudi Arabia Fireball Network (KSAFN). left: Fireball observation coverage in the Arabian peninsula as of January 2020. Each red dot corresponds to an observatory site. See Sec. \ref{sec:gfo_collab} for explanation on coverage area.
		right: Fireball observed from Alshaqiq station in the Ha'il region on Jan 31, 2020.}
		\label{fig:fig_Middle_East}
	\end{figure*}

Current partner networks are listed in Tab. \ref{table:subnets}.
Ground coverage for the GFO is plotted on maps for each part of the world in Figs. \ref{fig:aus_fig}- \ref{fig:northamerica_fig}- \ref{fig:morocco_fig}-\ref{fig:fig_UK}-\ref{fig:fig_Middle_East}.
To define coverage of a given area, we distinguish two criteria.
One for \textit{meteorite recovery}: based on the published details of previous falls in the literature (\citet{2015aste.book..257B} and references therein) and limits on precision of astrometric measurements close to horizon. We estimate that at least one camera needs to be closer than 130\,km to the meteoroid ground track, and a second viewpoint within 300\,km, in order to reliably calculate meteorite fall positions.
We also distinguish the sampling area of \textit{fireball orbits}, for which we relax the previous criterion to having two viewpoints within 300\,km.

With all networks combined, the Global Fireball Observatory collaboration covers 0.6\% of the Earth's surface for meteorite recovery (dark purple in Figs. \ref{fig:aus_fig}-\ref{fig:northamerica_fig}-\ref{fig:morocco_fig}-\ref{fig:fig_UK}-\ref{fig:fig_Middle_East}), and 1\% of Earth for orbital coverage (light purple).
In the early 2020s, the aim will be to cover 2\% of Earth for meteorite recovery, and 3\% for orbital sampling.

\begin{table*}
	\caption{Partner networks within the Global Fireball Observatory collaboration.}
	\label{table:subnets} 
		\begin{tabular}{| p{6.5cm} | p{4cm} | p{4.8cm} |}
			\hline
			\textbf{Network name} & \textbf{Region/Country} & \textbf{Managing institutions} \\
			\hline \hline
			Desert Fireball Network (DFN) & Western and South \textbf{Australia} - Fig. \ref{fig:aus_fig} & Curtin University \\
			\hline
			NASA Meteorite Tracking and Recovery Network & California and Nevada, \textbf{USA} - Fig. \ref{fig:northamerica_fig} & SETI Institute, NASA Ames Research Center\\
			\hline
			Moroccan Observatory for Fireball Detections (MOFID)  & \textbf{Morocco}  - Fig. \ref{fig:morocco_fig} & Hassan II University of Casablanca, Ouka\"{\i}meden Observatory\\
			\hline
			Meteorite Observation and Recovery Project 2.0 (MORP 2.0) & Alberta and Saskatchewan, \textbf{Canada} - Fig. \ref{fig:northamerica_fig} &  University of Alberta \\
			\hline
			Southern Ontario Meteor Network (SOMN)  & South-Western Ontario, \textbf{Canada} - Fig. \ref{fig:northamerica_fig} & University of Western Ontario \\
			\hline
			UK Fireball Network (UKFN) & \textbf{United Kingdom} - Fig. \ref{fig:fig_UK} & Imperial College London, University of Glasgow, University of Cambridge\\ 
			\hline
			TBD  & Australian Capital Territory, & Australian National University \\
			 &New South Wales - Fig. \ref{fig:aus_fig}& Macquarie University, Australia\\
			\hline
			TBD & Victoria, \textbf{Australia}	& Monash University\\
			\hline
			Kingdom of Saudi Arabia Fireball Network (KSAFN)   & \textbf{Kingdom of Saudi Arabia} - Fig. \ref{fig:fig_Middle_East} & National Center for Astronomy, King Abdulaziz City for Science and Technology \\
			\hline
			Arizona Fireball Network (AZFN)  & Arizona, \textbf{USA} - Fig. \ref{fig:northamerica_fig} & University of Arizona \\
			\hline
			TBD  & Texas, \textbf{USA} & NASA Johnson Space Center \\
			\hline
			TBD  & Queensland, \textbf{Australia} - Fig. \ref{fig:aus_fig} & University of Southern Queensland \\
			\hline
			TBD   & Tierra del Fuego, \textbf{Argentina}  & NASA Goddard Space Flight Center, Estacion Astronomica Rio Grande \\
			\hline
	\end{tabular}
\end{table*}

\section{Methods}

\subsection{Observatory hardware}\label{sec:hardware}

The observatories employed in the GFO are built upon the engineering heritage of the observatories used by the Desert Fireball Network in Australia.
The DFN observatories were designed with a strong focus on reliability and autonomy, with the first digital prototypes assembled in 2013, iterating to a design that was eventually mass-produced in 2014-2015 and rolled out to cover about a third of Australia using approximately 50 stations \citep{2017ExA...tmp...19H}.
To move beyond a network operated by the DFN and make a global expansion via a "network-of-networks" possible, changes to the design were required.
The motivation for the update was to improve the manufacturability of the design (as a large number of observatories were required in a short period of time) and to improve the maintainability and usability of the observatories \citep{2018PhDT.......RMH}.
The update notably aimed to improve cooling and add in the capability to heat the observatory for cold weather operation, allowing the system to operate from over 50$\degr$C on hot days in the Australian outback down to nearly -50$\degr$C in the Canadian winter.
The standard definition analogue video camera was replaced by a higher resolution (2.3MP) digital model capable of capturing periodic long exposure calibration frames as well as detect bright daytime bolides.
Finally, weatherproof external connections were added to improve observatory connectivity and allow plug-and-play auxiliary instruments such as a radiometer \citep{2019arXiv190712807B}.

\begin{figure}
	\centering
	\includegraphics[width=0.5\linewidth]{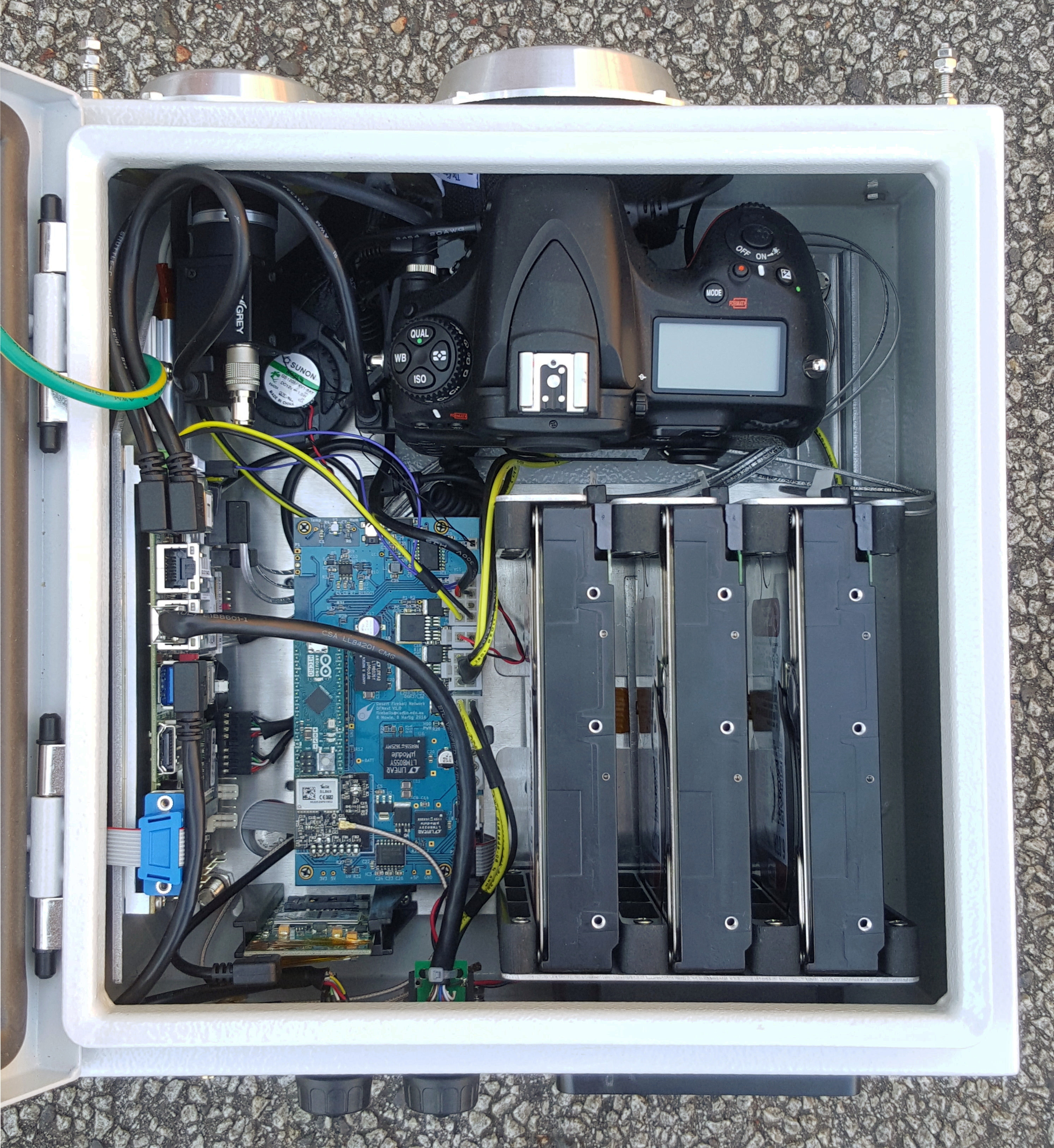}
	\caption{GFO fireball observatory showing (clockwise from bottom left) embedded PC, video camera, DSLR photographic camera, hard disc drives, observatory electronics board and 4G data modem, from \citep{2018PhDT.......RMH}}
	\label{fig:dfnext}
\end{figure}

Along with the hardware improvements, software improvements were also made including a move to a new pulse frequency encoding, resulting in improved fireball velocity data and the addition of periodic calibration frames unobscured by the operation of the liquid crystal shutter for improved astrometry.
The software and firmware of the improvements are backwards compatible with the previous observatories, which allows improvements made to be easily ported to older observatory models.
In addition, the new observatories also run a customised version of Freeture (the software designed to run the Fireball Recovery and Interplanetary Observation Network cameras \citep{2014pim4.conf...39A}) to handle the digital video observations.
The new revision of the observatory is shown in Fig. \ref{fig:dfnext} (cf. \cite{2017ExA...tmp...19H} Fig. 6).

\subsection{Data reduction} \label{sec:reduction_methods}

With the current number of cameras installed (${\sim}100$), over 6 TB of raw imagery is collected every night by the GFO.
An automated data reduction pipeline has been developed to quickly determine which images contain fireballs, and process these events to determine meteorite fall locations,
as well as calculate the meteoroids' dynamical origins (orbits).

Fireball event detection and corroboration between cameras are detailed by \citet{2019arXiv190911883T}, this part is mostly done on the embedded PCs on-board the observatories, while other steps of the pipeline are run on a central server. These include astrometric calibration of observational data, triangulation and analysis of fireball trajectories, orbit determination and darkflight modelling for meteorite recoveries.

Minute of arc astrometric precision is achieved with the method of \citet{2018PhDT.......HARD}, allowing reliable astrometric measurements of observed fireballs down to ${\sim}5\degr$ elevations for clear horizons, and ${\sim}10\degr$ in the case of more light polluted and/or partly obstructed skies.

The triangulation and trajectory analysis are performed using a variety of methods, as no single technique exists that can reliably determine the state vector of a large meteoroid throughout the trajectory with full uncertainty propagation.
We notably use the straight line trajectory determination method of \citet{1990BAICz..41..391B}, run in combination with various dynamical analysis techniques \citep{2015M&PS...50.1423S,2017AJ....153...87S,2019ApJ...885..115S}.
Some more modern approaches aim to derive the state vectors and physical parameters concurrently \citep{2019Icar..321..388S,2019arXiv191100816J}.

The pre-encounter orbit of a meteoroid is determined by numerical integration, as described by \citet{2019M&PS...54.2149J}.
On a typical fireball observed with our camera hardware, processed using the above methods, we get a pre-atmospheric speed formal uncertainty of ${\sim}60\,\mbox{m s}^{-1}$.
According to \citet{2018Icar..311..271G}, this precision on the meteoroid entry speed is generally sufficient to get accurate source region information when using Near-Earth Objects (NEO) population models like the one described by \citet{2018Icar..312..181G}.
The authors say that better precision would evidently lead to better results, but the difference is minor unless the speed and radiants measurements improve by orders of magnitude.

The meteorite search region is determined from the final conditions given by the triangulation and dynamic modelling stage, and an atmospheric model.
The atmospheric model is calculated following the methodology described by \citet{2018M&PS...53.2212D}.

From this model, the darkflight integrator interpolates wind speed, wind direction, pressure, temperature and relative humidity, at whatever position and time is required (available heights range up to ${\sim}$30\,km).
A number of virtual particles of varying mass, density, and shape are generated within the modelled uncertainty surrounding the meteoroid's final position and velocity along the observed trajectory.
These particles are then numerically integrated through their ballistic descent, under the influence of meteoroid ablation and atmospheric winds.
The numerical integration uses the 3D equations of the meteoroid's motion to realistically model the cosmic material until it reaches Earth's surface, producing a distribution of possible meteorite impact sites.
This Monte Carlo approach aims to encompass all uncertainties associated with the meteoroid state vector and physical properties, to derive probabilistic maps on the ground of where the meteorites likely landed (for an example see Fig. 10 of \citet{2018M&PS...53.2212D}).

\subsection{21st century meteorite searching techniques} 

Based on the knowledge gained during the meteoroid's initial bright flight phase (see Sec. \ref{sec:reduction_methods}), the position of the meteoroid must be numerically integrated through the last tens of kilometres of the atmosphere (the dark flight phase), carrying forward all the uncertainty on the state vector, physical characteristics (shape, mass, density, inner structure), and atmospheric conditions.
This process typically constrains the meteorite's fall location to an area on the order of a square kilometre for a favourable case, but up to several tens of square kilometres.
These large areas, combined with sometimes unfavourable searching terrains, can significantly inhibit meteorite recoveries.
Here we present some recent techniques that can help refine search areas.

\subsubsection{Weather Doppler radars}\label{sec:Doppler}

In some regions of the world, tight grids of weather Doppler radars have been set up to detect precipitation.
These can also be used to detect falling meteorites \citep{2014M&PS...49.1989F}.
As the radars scan very low on the horizon (down to $0.5\degr$), the altitudes at which the meteorites are detected are relatively low (sometimes down to a kilometre above the ground). This can lead to tightly constrained fall positions on the ground without necessarily taking the winds into account.

The detailed analysis of radar data have notably helped with some meteorite recoveries: Grimsby \citep{2011M&PS...46..339B}, Sutter's Mill \citep{2012Sci...338.1583J}, Creston \citep{2019M&PS...54..699J}, Dishchii'bikoh \citep{2019MNRAS.487.2307P}, and Hamburg \citep{2019M&PS...54.2027B}.
Although optical data were still used to determine the trajectory and orbit in these cases, the radar signatures were crucial to quickly locate the whereabouts of the meteorites on the ground, with a precision that exceeds what the observatory data alone would have been capable of achieving.

In areas with good radar coverage (mainly North America), we expect these data to simplify the recovery of GFO meteorites.
Also, with the GFO observatories now having daylight bolide detection capability (24h video capture), for these cases, we anticipate traditional meteorite recovery after darkflight integration to be even more difficult because of larger uncertainties; radar data will constitute important clues to help the recovery process.

\subsubsection{Use of small Uncrewed Aerial Vehicles (UAVs)}

With the hope that the ground search area has been reduced as much as possible, the recovery still depends on human vision and attention over long periods of time.
The deterioration of an individual's ability to identify signals or events over time has been documented as "vigilance decrement", it often becomes apparent after less than one hour of engaging in repetitive a task \citep{parasuraman1986vigilance,see1995meta}.
When considering that a typical search lasts for 8 hours per day, for sometimes more than 10 consecutive days, vigilance decrement becomes a serious problem.

To counter these issues, a dedicated team is working on automated meteorite searching techniques using a combination of \textit{robotic surveying} with small Uncrewed Aerial Vehicles (UAVs, also known as drones), and \textit{object recognition} using deep learning algorithms \citep{2017LPI....48.2528C,2019LPI....50.2426A}.

From a technology standpoint, the surveying part is relatively easy and is becoming cheap, thanks to the large commercial off-the-shelf development of small UAVs that come with easy to use control and surveying software.
Field tests show that one drone operator can reduce the total searching time by a factor of 10 compared to foot searching.

The real challenge lies with machine vision software.
The automated detection of objects in images using deep learning is a very active field of research \citep{NIPS2013_5207}.
These deep learning approaches typically require a lot of training data.
Acquiring a meteorite training dataset is a non-trivial problem, as meteorites vary significantly in size, shape, colour etc.
Nonetheless, initial tests show that, using state of the art deep learning technology, 95\% of the meteorites used for validation can be found.
In addition, it is equally important to minimise the number of false positive identifications which otherwise require manual evaluation by human researchers.

We expect the first live UAV searching tests on a real meteorite fall to be conducted in 2020 in Australia.
Once optimised, this technique will be used on several DFN fall sites that have not yet been searched because of a lack of person time, and eventually become a general tool for meteorite searching around the world.

\subsection{Other international efforts}

The Czech/European Fireball Network are covering a large fraction of central Europe, and have a long track record for recovering meteorites \citep{2015aste.book..257B}.
Their expertise in high-resolution long exposure camera systems contributed to shaping the DFN pathfinder project \citep{2012AuJES..59..177B}.

The Fireball Recovery and Interplanetary Observation Network (FRIPON) \citep{2015pimo.conf...37C} and their partners in Europe are well under way to covering a significant part of Western Europe.
They have chosen a radically different approach to observation hardware, with lower resolution observatories but on a much tighter grid, adding a level of reliability when poor weather conditions are present.
They developed reduction methods suitable for this different strategy \citep{2019A&A...627A..78J}.

Along with other smaller groups (see \citet{2019msme.book...90K} for a review), the global combined effort of fireball observation networks is going to create an unprecedented large web collecting 
centimetre to metre-scale objects impacting our planet.

\section{Likely GFO outcomes in the early 2020s}

\subsection{Meteorites}

There are $10^4$ sizeable meteorites ($>0.1$ kg) reaching the Earth's surface every year \citep{2006M&PS...41..607B}.
With 2\% of the Earth monitored by fireball observatories, assuming a conservative average of 75\% downtime because of daylight and cloudy conditions, the full GFO will observe ${\sim}50$ falls per year.
From basic statistics on the falls subset recorded in the Meteoritical Bulletin Database\footnote{\url{https://www.lpi.usra.edu/meteor/metbull.php}}, the main meteorite groups fall in these proportions:
\begin{itemize}
    \item Ordinary Chondrites: 70\% (8\% LL, 33\% L, 28\% H)
    \item Howardites, Eucrites, \& Diogenites (5\%)
    \item Carbonaceous chondrites (3.5\%)
    \item Irons (3.5\%)
\end{itemize}
The GFO will regularly observe many of these types of meteorites falling. Better still, it is statistically likely that representatives of almost every meteorite group will have been observed to fall over the course of 5 years of observation, including a Martian meteorite ($\simeq$1\% of falls).

Regardless of the type of material detected by the GFO, the advanced data reduction and concomitant potential for rapid recovery of freshly fallen meteorites from GFO data provide new opportunities to reduce the amount of time that meteorite specimens spend in the field, thus minimising terrestrial contamination and weathering \citep{2019SSRv..215...48M}.
Such events enable the application of best methods of curation in support of sample return, and build on lessons learned from previous falls (e.g. the organic-rich Tagish Lake meteorite \citep{2016M&PS...51..499H}).
In this way, meteorites recovered as a result of the GFO can be collected and transferred to curation facilities in such a manner as to preserve them against the oxidative, organic- and moisture-rich environment of the Earth's surface, and maximise their scientific return \citep{2019SSRv..215...48M}.

Here we present a selection of highlights of what questions the GFO might be able to answer over the next few years, given the recovery of a full suite of meteorite types.

\paragraph{Ordinary chondrites}
H chondrites are the group that has the most sample with orbits recovered, however there is still no consensus on what the parent body is \citep{2019M&PS...54.2027B}.
We hope that a large number of orbits for these objects can help pinpoint the various sources for this class of objects.
Although the sources for LL chondrites (broadly associated with the Flora family) and most of the shocked L chondrites (Gefion) are a little bit clearer, there are still some questions (see \citet{2020IAUGA..30....9J} for a review).

\paragraph{Iron meteorites}
Irons make up about 3.5\% of meteorite falls.
It would be fascinating to recover an iron meteorite with a well-defined orbit. Long Cosmic-Ray Exposure (CRE) ages are the norm for magmatic iron meteorites \citep{2006mess.book..829E}.
Defined peaks in CRE ages are clear in several groups, indicating discrete break-up events. Group III irons show a peak at around 650 Myr, and group IVs at around 400 Myr. On the other hand some irons have CRE ages exceeding 1 Gyr.
These extreme ages are certainly due in part to the strength and preferential survival of iron meteorites, but the fact that we see defined CRE age peaks indicates that their extreme ages may also be a product of an unusual orbital history.
The suggestion that the parent bodies for magmatic irons formed in the terrestrial planet region rather than at asteroidal distances, and that they were scattered into the main belt following interactions with planetary embryos \citep{2006Natur.439..821B}, may offer an explanation. It is possible that this could be resolved with high quality orbital data, allowing source region determination.

\paragraph{Carbonaceous chondrites}
Although carbonaceous chondrites with well-defined orbits have been recovered (C2-ungrouped \textit{Tagish Lake} \citep{2000Sci...290..320B}; CM2 \textit{Maribo} \citep{2019M&PS...54.1024B,2012M&PS...47...30H} and \textit{Sutter's Mill} \citep{2012Sci...338.1583J}), the dataset is currently too small to draw firm conclusions about the Solar System history of different groups, and how orbits relate to source regions and CRE ages.
Expanding this dataset is a headline priority for the GFO collaboration. The distribution of CRE ages varies widely between groups. The majority of CK and CV chondrites have ages in the range 8-30 Myr. CO chondrites show a diffuse peak at around 30 Myr.
But CM and CI chondrites are completely different. These meteorites have very short exposure ages.
CMs have a peak at 0.2 Myr, but some have ages of $<0.05$ Myr \citep{2009M&PSA..72.5358N}. The differences here likely reflect very different source regions for these groups.
It may be that CK/CV/CO chondrites are delivered from the main belt, while CI/CM come from a parent body on an Earth-crossing orbit.
A larger dataset of meteorites with orbits will allow us to determine their provenance.

\begin{figure}
	\centering
	\includegraphics[width=0.5\linewidth]{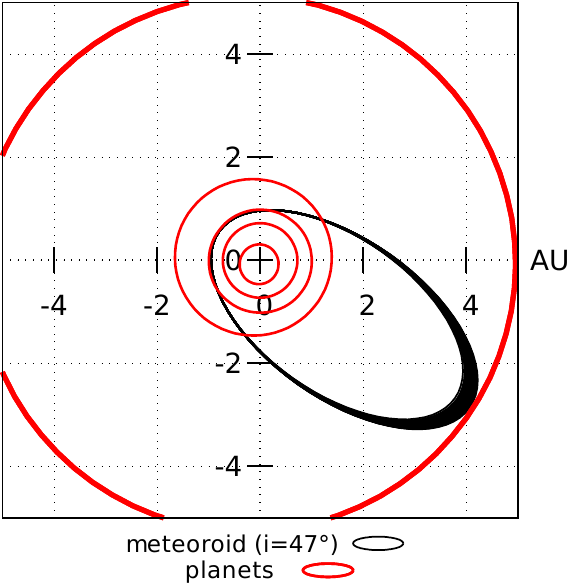}
	\caption{Ecliptic plot of the orbit of DFN meteorite dropping event \textit{DN190221\_03}. Its large inclination ($47\degr$) and its Tisserand criterion with Jupiter of 2.53 make this meteoroid well into the Jupiter Family comet region, beyond any connection with the main asteroid belt.}
	\label{fig:DN190221_03_MC_orbits}
\end{figure}

\paragraph{From a comet}
Comets are the most pristine material in the Solar System, containing a high-fidelity record of early Solar System processes, including the variety of stellar sources that contributed to our protoplanetary disk, and the earliest chemical processes that occurred within it \citep{2016A&A...592A..63D}.
They may have supplied Earth's water and organics.
The Stardust mission to recover $10^{-6}$ kg of Comet 81P/Wild 2 \citep{2006Sci...314.1711B} is testimony to the value placed on comets as witnesses of early Solar System processes.
The DFN recently observed a meteorite dropping fireball on a peculiar orbit (Fig. \ref{fig:DN190221_03_MC_orbits}).
This meteoroid is clearly dynamically de-coupled from the main belt. Although more work is required to determine its origin, it could be a fragment of a Jupiter family comet and maybe even be part of a meteor shower.
Unfortunately the estimated $\simeq0.1$ kg surviving mass fell into the ocean, annihilating chances of recovery, but at the same time proving that the endeavour of collecting meteorite samples goes well beyond a simple sampling of main belt material.

\paragraph{From meteor showers}
So far no meteorite has conclusively been associated with a meteor shower, but we know that this is possible, notably from the Geminids and Taurids streams \citep{2013M&PS...48..270B}.
There have been cases reported of Geminid meteoroids entering the atmosphere and convincingly leaving a non-zero mass \citep{spurny_2013_geminid,2013MNRAS.436.2818M}, a very large network such as the GFO should be able to observe these exceptional events on a more regular basis.

\subsection{Large dataset} 

Not every fireball observed leads to a meteorite, but the collation of all observed events contains precious clues about the near-Earth environment.
The distribution of orbits, strengths, and sizes all help build a picture of NEOs at the centimetre to metre-scale sizes.
So far the reference dataset remains the legacy work of \citet{1996M&PS...31..185H} on the MORP survey.

The DFN pathfinder project has already collected 575 $>0.1$ kg events over its initial 4 year survey, consistent with its 0.4\% Earth surveying area (using the size-frequency work of 
$10^5$ $>0.1$ kg impactors year$^{-1}$ Earth$^{-1}$ of \citet{2006M&PS...41..607B}).
While we wait for the meteorites with orbits to grow to statistically significant numbers, this large dataset of observed meteoroids is going to help refine the size-frequency distribution numbers of meteoroids at the centimetre to metre size ranges. This meteoroid orbit dataset could yield important insights on what might be happening at slightly larger asteroid sizes: there are still some questions about the size of the population of 10-50 m impactors \citep{2013Natur.503..238B}.
This size range has the potential to do damage on the ground, yet is poorly studied because of the lack of observations: the impacts on Earth are too infrequent, while the bodies are too small to be significantly observed by telescopes in sun-reflected light \citep{2019MNRAS.483.5166D}.
Constraining the impact flux on Earth can also help with the study of Mars, both for dating small areas / young surfaces, which have only accumulated small craters, and also for assessing the hazard to future human space exploration.
 This scale of impactors is difficult to detect on Mars because of the limits on resolution and coverage of current Mars imagery -- impact splotches can occasionally be detected in surface images taken by Mars orbiters, but only in some (dusty) regions of Mars and these features quickly fade.
 One goal of the current InSight mission is to detect small impacts using seismology to help address the observational bias and lack of good flux estimates \citep{2018SSRv..214..132D}.

Having a representative survey of the origin of asteroid material will also let us investigate further the lack of low perihelion NEOs proposed by \citet{2016Natur.530..303G}, and how this effect scales for small objects \citep{2018Icar..312..181G}.
Furthermore, a statistically significant number of orbits from suspected meteorite falls is going to help answer important questions.
Notably, can we reconcile the statistics on the number of meteorite falls and their classifications with the proportions of falls originating from various parts of the main asteroid belt?

\subsection{Probing the meteorite/asteroid link, synergy with NEO hunters}

A major goal in the study of small Solar System bodies is reconciling telescopic observations of asteroids and the study of their surface composition through reflectance spectra, with the meteorites analysed in the lab.
Other than through expensive sample return missions, the only way to get an irrefutable link between a spectral type of asteroid and a meteorite class is to have observed the meteorite progenitor before atmospheric impact.

Up to mid-2019, four asteroids have been detected before their confirmed impact on Earth: 2008 TC3 \citep{2009Natur.458..485J,2017Icar..294..218F}, 2014 AA \citep{2016Icar..274..327F}, and more recently 2018 LA and 2019 MO.
The two that impacted over land have led to the recovery of meteorites \citep{2019DDA....5020004F}.
These asteroids were detected by telescope facilities such as the Catalina Sky Survey (CSS), the Panoramic Survey Telescope and Rapid Response System (Pan-STARRS), and the Asteroid Terrestrial-impact Last Alert System (ATLAS).
Each have different and complementary NEO hunting strategies, with ATLAS covering the entire visible sky very rapidly at shallow depths, Pan-STARRS with a strong focus on a deep survey at opposition, and CSS somewhere in the middle \citep{2018PASP..130f4505T}.

With the Large Synoptic Survey Telescope (LSST) coming online in 2022 \citep{2019ApJ...873..111I}, the number of known asteroids is going to significantly increase, as is the number of NEOs and the number of imminent impactors.
The meteorite dropping objects observed by the GFO are typically decimetre to metre-scale, a size that will be detectable by LSST in the hours/days leading up to the impact, if the solar elongation approach is favourable.

\begin{figure}
\centering
\includegraphics[width=\linewidth]{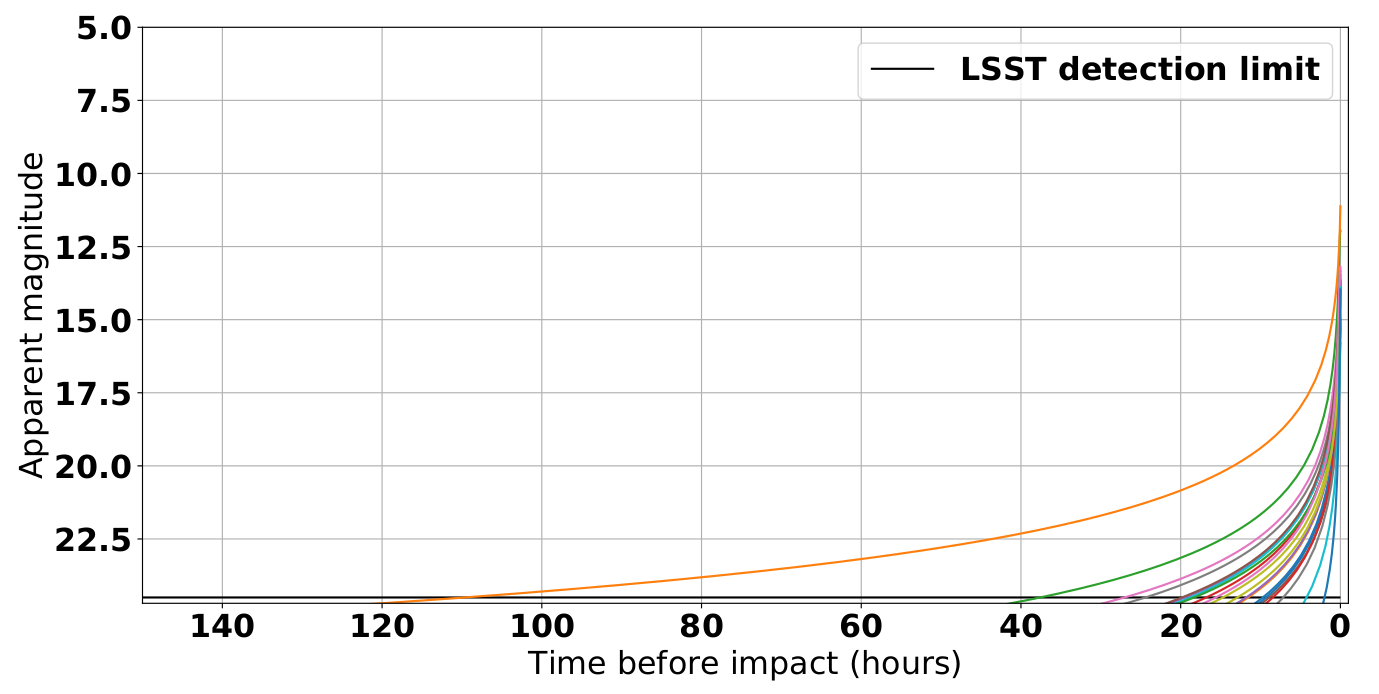}
\caption{Estimated pre-impact apparent magnitude of the 20 biggest objects observed by the DFN. The lower cut-off is set to the announced LSST point source limiting magnitude in a single visit (24.5 in V band). The meteoroid plotted in orange, visible for over 100 hours, is DN170630\_01, described by \citet{2019MNRAS.483.5166D}.}
\label{fig:DFN_LSST}
\end{figure}

\begin{figure}
	\centering
	\includegraphics[width=\linewidth]{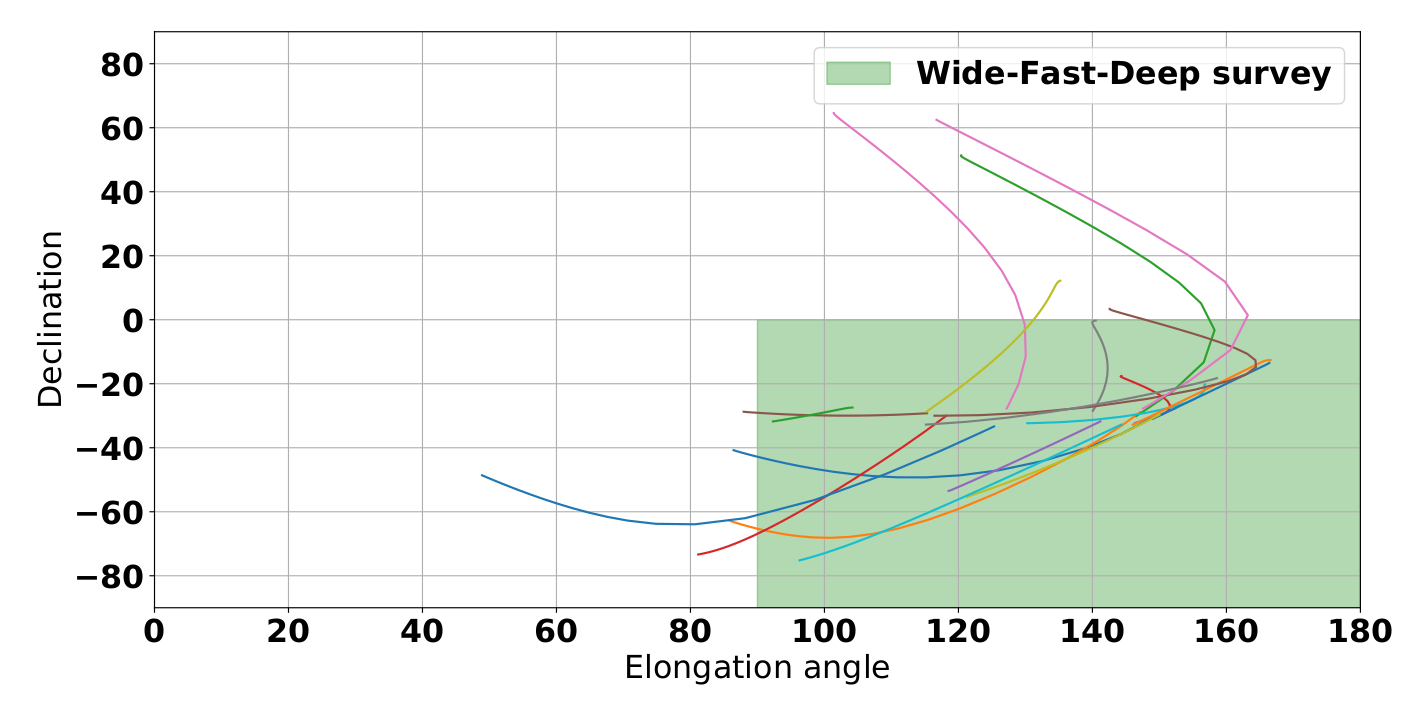}
	\caption{Pre-entry sky positions for the 20 biggest objects observed by the DFN. The paths are cut off once when the objects go fainter than apparent magnitude 24.5 (LSST limit). The green area corresponds to the so-called "Wide Fast Deep" survey of the LSST, with revisits expected every 3 days.}
	\label{fig:decl_elon}
\end{figure}

To get a better idea of what we can expect LSST to observe, we have carried out a small study, taking the 20 largest objects seen by the DFN pathfinder project during the first 4 years of science operations, and assuming LSST was online carrying its so-called "Wide Fast Deep" survey.
It is not clear at present what the observing strategy is exactly going to be, and the details of this strategy will have large implications for linking tracklets from fast moving objects together.
For simplicity here, the definition for a successful observation by LSST is if the telescope was to get a single picture of a specific object.
We also assume that LSST images the Southern sky at $>90\degr$ solar elongations once every 3 days.
We integrated backwards the positions of the meteoroid from the impact time, and generated ephemerides projected into (solar elongation, declination) coordinates, as these coordinates are easily relatable with the LSST survey.
Along with the ephemerides, we calculated the illumination of the targets and their brightness over time, assuming an S-type albedo (0.15) for all objects.
Most of our large objects are only visible in the last 10-20 hours before impact (Fig. \ref{fig:DFN_LSST}), and most have relatively large solar elongations (Fig. \ref{fig:decl_elon}).
This skew towards large solar elongations likely comes from the fact that the objects impacted at night time.
Also, most of the objects seem to spend their last hours in Southern declinations (Fig. \ref{fig:decl_elon}).
This observation bias is caused by the mostly Southern location of the DFN pathfinder project, it is not expected to scale with a Global Fireball Observatory.
We note that \textit{DN170630\_01}, a 1.3 m object observed in South Australia \citep{2019MNRAS.483.5166D}, would have remained visible to LSST for over 4 days before impact, and therefore would have had a very high chance of being detected.

Statistically speaking, when we convolve the observability of these 20 objects with the Wide-Fast-Deep survey of LSST, we get 3.6 objects detected by LSST.
We estimate the coverage of the Earth to have been ${\sim}0.4\%$ for the DFN pathfinder project over these 4 years.
Scaling this to the Global Fireball Observatory (2\% Earth) would give us ${\sim}2$ GFO meteoroids pre-detected every year by LSST.
It is not exactly clear how many of these would actually be flagged as new objects and sent as alerts by the LSST processing pipeline (see \citet{2017arXiv170506209C} for a review).
This alert process would be helpful to enable follow-up observations before impact, as was the case for \textit{2008 TC3}.

In all cases, the GFO/LSST synergy will not only give strong astrometric constraints to refine calculated orbits, but also help constrain other characteristic properties such as colours, rotations, and albedos, providing further insights into asteroids/meteorites links.

\subsection{Other uses of the data}

The GFO's ability to constantly monitor large areas of the sky can also be applied to transient astronomy.
Although rare, the brightest transient events are also the most interesting because they are easy to follow up spectroscopically.

These transients include optical counterparts of Gamma-Ray Bursts. In 2008, GRB 080319B achieved a brightness of V = 5.3 \citep{2008Natur.455..183R}, well within the magnitude range reachable by a single GFO exposure.

In the era of multi-messenger astronomy, quickly determining the location of gravitational waves is also an area where the GFO can contribute. Albeit not sensitive enough for this particular event, the DFN pathfinder project was notably the first optical observatory on-sky for the neutron star merger GW170817 \citep{2017ApJ...848L..12A,2017PASA...34...69A}.

The fact that GFO observatories are already recording imagery before the events happen is useful because some transients cannot be followed up by traditional methods.
For example, fast radio bursts experience a frequency-dependent time shift, which would make any emission in the optical arrive before any radio signals are detected \citep{2019ApJ...872L..19M,2019Sci...365..565B}.

\section*{Acknowledgements}

This research is supported by the Australian Research Council through the Linkage Infrastructure, Equipment and Facilities program (LE170100106).
The DFN receives institutional support from Curtin University, and uses the computing facilities of the Pawsey supercomputing centre. The team would like to thank the people hosting the observatories.

The NASA Tracking and Recovery Network is funded by NASA grant 80 NSSC18K08. PJ acknowledges logistic support from NASA's SERVII program.

The UKFN would like to thank a number of people helping with the project: Ian McMullan, Adam Suttle, Pierre-Etienne Martin, Cameron Floyd, Áine O' Brien, Sammy Griffin, Charlotte Slaymark, Annemarie E. Pickersgill, Peter Chung, Joshua F. Einsle, Mohammad Ali Salik.

The data reduction pipeline makes intensive use of Astropy, a community-developed core Python package for Astronomy \citep{2013A&A...558A..33A}.
\printcredits

\bibliographystyle{mnras}

\bibliography{research}

\begin{thebibliography}{}
\makeatletter
\relax
\def\mn@urlcharsother{\let\do\@makeother \do\$\do\&\do\#\do\^\do\_\do\%\do\~}
\def\mn@doi{\begingroup\mn@urlcharsother \@ifnextchar [ {\mn@doi@}
  {\mn@doi@[]}}
\def\mn@doi@[#1]#2{\def\@tempa{#1}\ifx\@tempa\@empty \href
  {http://dx.doi.org/#2} {doi:#2}\else \href {http://dx.doi.org/#2} {#1}\fi
  \endgroup}
\def\mn@eprint#1#2{\mn@eprint@#1:#2::\@nil}
\def\mn@eprint@arXiv#1{\href {http://arxiv.org/abs/#1} {{\tt arXiv:#1}}}
\def\mn@eprint@dblp#1{\href {http://dblp.uni-trier.de/rec/bibtex/#1.xml}
  {dblp:#1}}
\def\mn@eprint@#1:#2:#3:#4\@nil{\def\@tempa {#1}\def\@tempb {#2}\def\@tempc
  {#3}\ifx \@tempc \@empty \let \@tempc \@tempb \let \@tempb \@tempa \fi \ifx
  \@tempb \@empty \def\@tempb {arXiv}\fi \@ifundefined
  {mn@eprint@\@tempb}{\@tempb:\@tempc}{\expandafter \expandafter \csname
  mn@eprint@\@tempb\endcsname \expandafter{\@tempc}}}

\bibitem[\protect\citeauthoryear{{Anderson}, {Bland}, {Towner}  \&
  {Paxman}}{{Anderson} et~al.}{2019}]{2019LPI....50.2426A}
{Anderson} S.~L.,  {Bland} P.~A.,  {Towner} M.~C.,   {Paxman} J.~P.,  2019, in
  Lunar and Planetary Science Conference. Lunar and Planetary Science
  Conference.
p.~2426

\bibitem[\protect\citeauthoryear{{Andreoni} et~al.,}{{Andreoni}
  et~al.}{2017}]{2017PASA...34...69A}
{Andreoni} I.,  et~al., 2017, \mn@doi [Publications of the Astronomical Society
  of Australia] {10.1017/pasa.2017.65}, \href
  {https://ui.adsabs.harvard.edu/#abs/2017PASA...34...69A} {34, e069}

\bibitem[\protect\citeauthoryear{{Astropy Collaboration} et~al.,}{{Astropy
  Collaboration} et~al.}{2013}]{2013A&A...558A..33A}
{Astropy Collaboration} et~al., 2013, \mn@doi [\aap]
  {10.1051/0004-6361/201322068}, \href
  {http://adsabs.harvard.edu/abs/2013A%26A...558A..33A} {558, A33}

\bibitem[\protect\citeauthoryear{{Audureau} et~al.,}{{Audureau}
  et~al.}{2014}]{2014pim4.conf...39A}
{Audureau} Y.,  et~al., 2014, in {Rault} J.-L.,  {Roggemans} P.,  eds,
  Proceedings of the International Meteor Conference, Giron, France, 18-21
  September 2014. pp 39--41

\bibitem[\protect\citeauthoryear{{Bannister} et~al.,}{{Bannister}
  et~al.}{2019}]{2019Sci...365..565B}
{Bannister} K.~W.,  et~al., 2019, \mn@doi [Science] {10.1126/science.aaw5903},
  \href {https://ui.adsabs.harvard.edu/abs/2019Sci...365..565B} {365, 565}

\bibitem[\protect\citeauthoryear{{Bland} \& {Artemieva}}{{Bland} \&
  {Artemieva}}{2006}]{2006M&PS...41..607B}
{Bland} P.~A.,  {Artemieva} N.~A.,  2006, \mn@doi [Meteoritics and Planetary
  Science] {10.1111/j.1945-5100.2006.tb00485.x}, \href
  {https://ui.adsabs.harvard.edu/#abs/2006M&PS...41..607B} {41, 607}

\bibitem[\protect\citeauthoryear{{Bland} et~al.,}{{Bland}
  et~al.}{2012}]{2012AuJES..59..177B}
{Bland} P.~A.,  et~al., 2012, \mn@doi [Australian Journal of Earth Sciences]
  {10.1080/08120099.2011.595428}, \href
  {http://adsabs.harvard.edu/abs/2012AuJES..59..177B} {59, 177}

\bibitem[\protect\citeauthoryear{{Borovi{\v c}ka}}{{Borovi{\v
  c}ka}}{1990}]{1990BAICz..41..391B}
{Borovi{\v c}ka} J.,  1990, Bulletin of the Astronomical Institutes of
  Czechoslovakia, \href {http://adsabs.harvard.edu/abs/1990BAICz..41..391B}
  {41, 391}

\bibitem[\protect\citeauthoryear{{Borovi{\v c}ka}, {Spurn{\'y}}  \&
  {Brown}}{{Borovi{\v c}ka} et~al.}{2015}]{2015aste.book..257B}
{Borovi{\v c}ka} J.,  {Spurn{\'y}} P.,   {Brown} P.,  2015, {Small Near-Earth
  Asteroids as a Source of Meteorites}.
University of Arizona Press, pp 257--280,
  \mn@doi{10.2458/azu_uapress_9780816532131-ch014}

\bibitem[\protect\citeauthoryear{{Borovi{\v{c}}ka}, {Popova}  \&
  {Spurn{\'y}}}{{Borovi{\v{c}}ka} et~al.}{2019}]{2019M&PS...54.1024B}
{Borovi{\v{c}}ka} J.,  {Popova} O.,   {Spurn{\'y}} P.,  2019, \mn@doi
  [Meteoritics and Planetary Science] {10.1111/maps.13259}, \href
  {https://ui.adsabs.harvard.edu/abs/2019M&PS...54.1024B} {54, 1024}

\bibitem[\protect\citeauthoryear{{Bottke}, {Nesvorn{\'y}}, {Grimm},
  {Morbidelli}  \& {O'Brien}}{{Bottke} et~al.}{2006}]{2006Natur.439..821B}
{Bottke} W.~F.,  {Nesvorn{\'y}} D.,  {Grimm} R.~E.,  {Morbidelli} A.,
  {O'Brien} D.~P.,  2006, \mn@doi [\nat] {10.1038/nature04536}, \href
  {https://ui.adsabs.harvard.edu/abs/2006Natur.439..821B} {439, 821}

\bibitem[\protect\citeauthoryear{{Brown} et~al.,}{{Brown}
  et~al.}{2000}]{2000Sci...290..320B}
{Brown} P.~G.,  et~al., 2000, \mn@doi [Science] {10.1126/science.290.5490.320},
  \href {http://adsabs.harvard.edu/abs/2000Sci...290..320B} {290, 320}

\bibitem[\protect\citeauthoryear{{Brown} et~al.,}{{Brown}
  et~al.}{2011}]{2011M&PS...46..339B}
{Brown} P.,  et~al., 2011, \mn@doi [Meteoritics and Planetary Science]
  {10.1111/j.1945-5100.2010.01167.x}, \href
  {http://adsabs.harvard.edu/abs/2011M%26PS...46..339B} {46, 339}

\bibitem[\protect\citeauthoryear{{Brown}, {Marchenko}, {Moser}, {Weryk}  \&
  {Cooke}}{{Brown} et~al.}{2013a}]{2013M&PS...48..270B}
{Brown} P.,  {Marchenko} V.,  {Moser} D.~E.,  {Weryk} R.,   {Cooke} W.,  2013a,
  \mn@doi [Meteoritics and Planetary Science] {10.1111/maps.12055}, \href
  {http://adsabs.harvard.edu/abs/2013M%26PS...48..270B} {48, 270}

\bibitem[\protect\citeauthoryear{{Brown} et~al.,}{{Brown}
  et~al.}{2013b}]{2013Natur.503..238B}
{Brown} P.~G.,  et~al., 2013b, \mn@doi [\nat] {10.1038/nature12741}, \href
  {http://adsabs.harvard.edu/abs/2013Natur.503..238B} {503, 238}

\bibitem[\protect\citeauthoryear{{Brown} et~al.,}{{Brown}
  et~al.}{2019}]{2019M&PS...54.2027B}
{Brown} P.~G.,  et~al., 2019, \mn@doi [Meteoritics and Planetary Science]
  {10.1111/maps.13368}, \href
  {https://ui.adsabs.harvard.edu/abs/2019M&PS...54.2027B} {54, 2027}

\bibitem[\protect\citeauthoryear{{Brownlee} et~al.,}{{Brownlee}
  et~al.}{2006}]{2006Sci...314.1711B}
{Brownlee} D.,  et~al., 2006, \mn@doi [Science] {10.1126/science.1135840},
  \href {https://ui.adsabs.harvard.edu/abs/2006Sci...314.1711B} {314, 1711}

\bibitem[\protect\citeauthoryear{{Buchan}, {Howie}, {Paxman}  \&
  {Devillepoix}}{{Buchan} et~al.}{2019}]{2019arXiv190712807B}
{Buchan} S. R.~G.,  {Howie} R.~M.,  {Paxman} J.,   {Devillepoix} H. A.~R.,
  2019, arXiv e-prints, \href
  {https://ui.adsabs.harvard.edu/abs/2019arXiv190712807B} {p. arXiv:1907.12807}

\bibitem[\protect\citeauthoryear{{Chennaoui Aoudjehane}}{{Chennaoui
  Aoudjehane}}{2016}]{2016LPICo1921.6119C}
{Chennaoui Aoudjehane} H.,  2016, in 79th Annual Meeting of the Meteoritical
  Society. p.~6119

\bibitem[\protect\citeauthoryear{{Chennaoui Aoudjehane} \& {Agee}}{{Chennaoui
  Aoudjehane} \& {Agee}}{2019}]{2019LPICo2157.6297C}
{Chennaoui Aoudjehane} H.,  {Agee} C.~B.,  2019, LPI Contributions, \href
  {https://ui.adsabs.harvard.edu/abs/2019LPICo2157.6297C} {2157, 6297}

\bibitem[\protect\citeauthoryear{{Chennaoui Aoudjehane} et~al.,}{{Chennaoui
  Aoudjehane} et~al.}{2012}]{2012Sci...338..785A}
{Chennaoui Aoudjehane} H.,  et~al., 2012, \mn@doi [Science]
  {10.1126/science.1224514}, \href
  {https://ui.adsabs.harvard.edu/abs/2012Sci...338..785A} {338, 785}

\bibitem[\protect\citeauthoryear{{Chesley} \& {Veres}}{{Chesley} \&
  {Veres}}{2017}]{2017arXiv170506209C}
{Chesley} S.~R.,  {Veres} P.,  2017, arXiv e-prints, \href
  {https://ui.adsabs.harvard.edu/abs/2017arXiv170506209C} {p. arXiv:1705.06209}

\bibitem[\protect\citeauthoryear{{Citron}, {Shah}, {Sinha}, {Watkins}  \&
  {Jenniskens}}{{Citron} et~al.}{2017}]{2017LPI....48.2528C}
{Citron} R.~I.,  {Shah} A.,  {Sinha} S.,  {Watkins} C.,   {Jenniskens} P.,
  2017, in Lunar and Planetary Science Conference. p.~2528

\bibitem[\protect\citeauthoryear{{Colas} et~al.,}{{Colas}
  et~al.}{2015}]{2015pimo.conf...37C}
{Colas} F.,  et~al., 2015, in {Rault} J.-L.,  {Roggemans} P.,  eds,
  International Meteor Conference Mistelbach, Austria. pp 37--40

\bibitem[\protect\citeauthoryear{{Daubar} et~al.,}{{Daubar}
  et~al.}{2018}]{2018SSRv..214..132D}
{Daubar} I.,  et~al., 2018, \mn@doi [\ssr] {10.1007/s11214-018-0562-x}, \href
  {https://ui.adsabs.harvard.edu/abs/2018SSRv..214..132D} {214, 132}

\bibitem[\protect\citeauthoryear{{Davidsson} et~al.,}{{Davidsson}
  et~al.}{2016}]{2016A&A...592A..63D}
{Davidsson} B.~J.~R.,  et~al., 2016, \mn@doi [\aap]
  {10.1051/0004-6361/201526968}, \href
  {https://ui.adsabs.harvard.edu/abs/2016A&A...592A..63D} {592, A63}

\bibitem[\protect\citeauthoryear{{Devillepoix}}{{Devillepoix}}{2018}]{2018PhDT.......HARD}
{Devillepoix} H.~A.~R.,  2018, PhD thesis, School of Earth and Planetary
  Sciences, Curtin University, \url
  {https://espace.curtin.edu.au/handle/20.500.11937/76001}

\bibitem[\protect\citeauthoryear{{Devillepoix} et~al.,}{{Devillepoix}
  et~al.}{2018}]{2018M&PS...53.2212D}
{Devillepoix} H. A.~R.,  et~al., 2018, \mn@doi [Meteoritics and Planetary
  Science] {10.1111/maps.13142}, \href
  {https://ui.adsabs.harvard.edu/#abs/2018M&PS...53.2212D} {53, 2212}

\bibitem[\protect\citeauthoryear{{Devillepoix} et~al.,}{{Devillepoix}
  et~al.}{2019}]{2019MNRAS.483.5166D}
{Devillepoix} H.~A.~R.,  et~al., 2019, \mn@doi [\mnras]
  {10.1093/mnras/sty3442}, \href
  {https://ui.adsabs.harvard.edu/\#abs/2019MNRAS.483.5166D} {483, 5166}

\bibitem[\protect\citeauthoryear{{Eugster}, {Herzog}, {Marti}  \&
  {Caffee}}{{Eugster} et~al.}{2006}]{2006mess.book..829E}
{Eugster} O.,  {Herzog} G.~F.,  {Marti} K.,   {Caffee} M.~W.,  2006,
  {Irradiation Records, Cosmic-Ray Exposure Ages, and Transfer Times of
  Meteorites}.
p.~829

\bibitem[\protect\citeauthoryear{{Farnocchia}, {Chesley}, {Brown}  \&
  {Chodas}}{{Farnocchia} et~al.}{2016}]{2016Icar..274..327F}
{Farnocchia} D.,  {Chesley} S.~R.,  {Brown} P.~G.,   {Chodas} P.~W.,  2016,
  \mn@doi [\icarus] {10.1016/j.icarus.2016.02.056}, \href
  {http://adsabs.harvard.edu/abs/2016Icar..274..327F} {274, 327}

\bibitem[\protect\citeauthoryear{{Farnocchia}, {Jenniskens}, {Robertson},
  {Chesley}, {Dimare}  \& {Chodas}}{{Farnocchia}
  et~al.}{2017}]{2017Icar..294..218F}
{Farnocchia} D.,  {Jenniskens} P.,  {Robertson} D.~K.,  {Chesley} S.~R.,
  {Dimare} L.,   {Chodas} P.~W.,  2017, \mn@doi [\icarus]
  {10.1016/j.icarus.2017.03.007}, \href
  {http://adsabs.harvard.edu/abs/2017Icar..294..218F} {294, 218}

\bibitem[\protect\citeauthoryear{{Farnocchia}, {Chesley}, {Chodas},
  {Christensen}, {Kowalski}, {Brown}  \& {Jenniskens}}{{Farnocchia}
  et~al.}{2019}]{2019DDA....5020004F}
{Farnocchia} D.,  {Chesley} S.~R.,  {Chodas} P.~W.,  {Christensen} E.,
  {Kowalski} R.~A.,  {Brown} P.~G.,   {Jenniskens} P.,  2019, in AAS/Division
  of Dynamical Astronomy Meeting. p. 200.04

\bibitem[\protect\citeauthoryear{{Fries}, {Le Corre}, {Hankey}, {Fries},
  {Matson}, {Schaefer}  \& {Reddy}}{{Fries} et~al.}{2014}]{2014M&PS...49.1989F}
{Fries} M.,  {Le Corre} L.,  {Hankey} M.,  {Fries} J.,  {Matson} R.,
  {Schaefer} J.,   {Reddy} V.,  2014, \mn@doi [Meteoritics and Planetary
  Science] {10.1111/maps.12249}, \href
  {http://adsabs.harvard.edu/abs/2014M%26PS...49.1989F} {49, 1989}

\bibitem[\protect\citeauthoryear{{Granvik} \& {Brown}}{{Granvik} \&
  {Brown}}{2018}]{2018Icar..311..271G}
{Granvik} M.,  {Brown} P.,  2018, \mn@doi [\icarus]
  {10.1016/j.icarus.2018.04.012}, \href
  {https://ui.adsabs.harvard.edu/#abs/2018Icar..311..271G} {311, 271}

\bibitem[\protect\citeauthoryear{{Granvik} et~al.,}{{Granvik}
  et~al.}{2016}]{2016Natur.530..303G}
{Granvik} M.,  et~al., 2016, \mn@doi [\nat] {10.1038/nature16934}, \href
  {https://ui.adsabs.harvard.edu/#abs/2016Natur.530..303G} {530, 303}

\bibitem[\protect\citeauthoryear{{Granvik} et~al.,}{{Granvik}
  et~al.}{2018}]{2018Icar..312..181G}
{Granvik} M.,  et~al., 2018, \mn@doi [\icarus] {10.1016/j.icarus.2018.04.018},
  \href {https://ui.adsabs.harvard.edu/#abs/2018Icar..312..181G} {312, 181}

\bibitem[\protect\citeauthoryear{{Haack} et~al.,}{{Haack}
  et~al.}{2012}]{2012M&PS...47...30H}
{Haack} H.,  et~al., 2012, \mn@doi [Meteoritics and Planetary Science]
  {10.1111/j.1945-5100.2011.01311.x}, \href
  {https://ui.adsabs.harvard.edu/abs/2012M&PS...47...30H} {47, 30}

\bibitem[\protect\citeauthoryear{{Halliday}, {Blackwell}  \&
  {Griffin}}{{Halliday} et~al.}{1989}]{1989Metic..24...65H}
{Halliday} I.,  {Blackwell} A.~T.,   {Griffin} A.~A.,  1989, \mn@doi
  [Meteoritics] {10.1111/j.1945-5100.1989.tb00946.x}, \href
  {https://ui.adsabs.harvard.edu/abs/1989Metic..24...65H} {24, 65}

\bibitem[\protect\citeauthoryear{{Halliday}, {Griffin}  \&
  {Blackwell}}{{Halliday} et~al.}{1996}]{1996M&PS...31..185H}
{Halliday} I.,  {Griffin} A.~A.,   {Blackwell} A.~T.,  1996, \mn@doi
  [Meteoritics and Planetary Science] {10.1111/j.1945-5100.1996.tb02014.x},
  \href {http://adsabs.harvard.edu/abs/1996M%26PS...31..185H} {31, 185}

\bibitem[\protect\citeauthoryear{{Herd}, {Hilts}, {Skelhorne}  \&
  {Simkus}}{{Herd} et~al.}{2016}]{2016M&PS...51..499H}
{Herd} C. D.~K.,  {Hilts} R.~W.,  {Skelhorne} A.~W.,   {Simkus} D.~N.,  2016,
  \mn@doi [Meteoritics and Planetary Science] {10.1111/maps.12603}, \href
  {https://ui.adsabs.harvard.edu/abs/2016M&PS...51..499H} {51, 499}

\bibitem[\protect\citeauthoryear{{Howie}}{{Howie}}{2019}]{2018PhDT.......RMH}
{Howie} R.~M.,  2019, PhD thesis, Mechanical Engineering, Curtin University,
  \url {http://hdl.handle.net/20.500.11937/75046}

\bibitem[\protect\citeauthoryear{{Howie}, {Paxman}, {Bland}, {Towner},
  {Cup\'{a}k}, {Sansom}  \& {Devillepoix}}{{Howie}
  et~al.}{2017}]{2017ExA...tmp...19H}
{Howie} R.~M.,  {Paxman} J.,  {Bland} P.~A.,  {Towner} M.~C.,  {Cup\'{a}k} M.,
  {Sansom} E.~K.,   {Devillepoix} H.~A.~R.,  2017, \mn@doi [Experimental
  Astronomy] {10.1007/s10686-017-9532-7}, \href
  {http://adsabs.harvard.edu/abs/2017ExA...tmp...19H} {}

\bibitem[\protect\citeauthoryear{{Ivezi{\'c}} et~al.,}{{Ivezi{\'c}}
  et~al.}{2019}]{2019ApJ...873..111I}
{Ivezi{\'c}} {\v{Z}}.,  et~al., 2019, \mn@doi [\apj]
  {10.3847/1538-4357/ab042c}, \href
  {https://ui.adsabs.harvard.edu/abs/2019ApJ...873..111I} {873, 111}

\bibitem[\protect\citeauthoryear{{Jansen-Sturgeon}, {Sansom}, {Devillepoix},
  {Bland}, {Towner}, {Howie}  \& {Hartig}}{{Jansen-Sturgeon}
  et~al.}{2019a}]{2019arXiv191100816J}
{Jansen-Sturgeon} T.,  {Sansom} E.~K.,  {Devillepoix} H. A.~R.,  {Bland} P.~A.,
   {Towner} M.~C.,  {Howie} R.~M.,   {Hartig} B. A.~D.,  2019a, arXiv e-prints,
  \href {https://ui.adsabs.harvard.edu/abs/2019arXiv191100816J} {p.
  arXiv:1911.00816}

\bibitem[\protect\citeauthoryear{{Jansen-Sturgeon}, {Sansom}  \&
  {Bland}}{{Jansen-Sturgeon} et~al.}{2019b}]{2019M&PS...54.2149J}
{Jansen-Sturgeon} T.,  {Sansom} E.~K.,   {Bland} P.~A.,  2019b, \mn@doi
  [Meteoritics and Planetary Science] {10.1111/maps.13376}, \href
  {https://ui.adsabs.harvard.edu/abs/2019M&PS...54.2149J} {54, 2149}

\bibitem[\protect\citeauthoryear{{Jeanne} et~al.,}{{Jeanne}
  et~al.}{2019}]{2019A&A...627A..78J}
{Jeanne} S.,  et~al., 2019, \mn@doi [\aap] {10.1051/0004-6361/201834990}, \href
  {https://ui.adsabs.harvard.edu/abs/2019A&A...627A..78J} {627, A78}

\bibitem[\protect\citeauthoryear{{Jenniskens}}{{Jenniskens}}{2014}]{2014me13.conf...57J}
{Jenniskens} P.,  2014, Meteoroids 2013, \href
  {http://adsabs.harvard.edu/abs/2014me13.conf...57J} {}

\bibitem[\protect\citeauthoryear{{Jenniskens}}{{Jenniskens}}{2020}]{2020IAUGA..30....9J}
{Jenniskens} P.,  2020, in IAU General Assembly. pp 9--12,
  \mn@doi{10.1017/S1743921319003235}

\bibitem[\protect\citeauthoryear{{Jenniskens} et~al.,}{{Jenniskens}
  et~al.}{2009}]{2009Natur.458..485J}
{Jenniskens} P.,  et~al., 2009, \mn@doi [\nat] {10.1038/nature07920}, \href
  {http://adsabs.harvard.edu/abs/2009Natur.458..485J} {458, 485}

\bibitem[\protect\citeauthoryear{{Jenniskens} et~al.,}{{Jenniskens}
  et~al.}{2012}]{2012Sci...338.1583J}
{Jenniskens} P.,  et~al., 2012, \mn@doi [Science] {10.1126/science.1227163},
  \href {http://adsabs.harvard.edu/abs/2012Sci...338.1583J} {338, 1583}

\bibitem[\protect\citeauthoryear{{Jenniskens} et~al.,}{{Jenniskens}
  et~al.}{2019}]{2019M&PS...54..699J}
{Jenniskens} P.,  et~al., 2019, \mn@doi [Meteoritics and Planetary Science]
  {10.1111/maps.13235}, \href
  {https://ui.adsabs.harvard.edu/abs/2019M&PS...54..699J} {54, 699}

\bibitem[\protect\citeauthoryear{{Koten}, {Rendtel}, {Shrben{\'y}}, {Gural},
  {Borovi{\v{c}}ka}  \& {Kozak}}{{Koten} et~al.}{2019}]{2019msme.book...90K}
{Koten} P.,  {Rendtel} J.,  {Shrben{\'y}} L.,  {Gural} P.,  {Borovi{\v{c}}ka}
  J.,   {Kozak} P.,  2019, {Meteors and Meteor Showers as Observed by Optical
  Techniques}.
p.~90, \mn@doi{Here DOI}

\bibitem[\protect\citeauthoryear{{LIGO Scientific Collaboration} et~al.,}{{LIGO
  Scientific Collaboration} et~al.}{2017}]{2017ApJ...848L..12A}
{LIGO Scientific Collaboration} et~al., 2017, \mn@doi [\apj]
  {10.3847/2041-8213/aa91c9}, \href
  {https://ui.adsabs.harvard.edu/#abs/2017ApJ...848L..12A} {848, L12}

\bibitem[\protect\citeauthoryear{{Macquart}, {Shannon}, {Bannister}, {James},
  {Ekers}  \& {Bunton}}{{Macquart} et~al.}{2019}]{2019ApJ...872L..19M}
{Macquart} J.~P.,  {Shannon} R.~M.,  {Bannister} K.~W.,  {James} C.~W.,
  {Ekers} R.~D.,   {Bunton} J.~D.,  2019, \mn@doi [\apjl]
  {10.3847/2041-8213/ab03d6}, \href
  {https://ui.adsabs.harvard.edu/abs/2019ApJ...872L..19M} {872, L19}

\bibitem[\protect\citeauthoryear{{Madiedo}, {Trigo-Rodr{\'\i}guez},
  {Castro-Tirado}, {Ortiz}  \& {Cabrera-Ca{\~n}o}}{{Madiedo}
  et~al.}{2013}]{2013MNRAS.436.2818M}
{Madiedo} J.~M.,  {Trigo-Rodr{\'\i}guez} J.~M.,  {Castro-Tirado} A.~J.,
  {Ortiz} J.~L.,   {Cabrera-Ca{\~n}o} J.,  2013, \mn@doi [\mnras]
  {10.1093/mnras/stt1777}, \href
  {https://ui.adsabs.harvard.edu/abs/2013MNRAS.436.2818M} {436, 2818}

\bibitem[\protect\citeauthoryear{{McCubbin} et~al.,}{{McCubbin}
  et~al.}{2019}]{2019SSRv..215...48M}
{McCubbin} F.~M.,  et~al., 2019, \mn@doi [\ssr] {10.1007/s11214-019-0615-9},
  \href {https://ui.adsabs.harvard.edu/abs/2019SSRv..215...48M} {215, 48}

\bibitem[\protect\citeauthoryear{{Nishiizumi} \& {Caffee}}{{Nishiizumi} \&
  {Caffee}}{2009}]{2009M&PSA..72.5358N}
{Nishiizumi} K.,  {Caffee} M.~W.,  2009, Meteoritics and Planetary Science
  Supplement, \href {https://ui.adsabs.harvard.edu/abs/2009M&PSA..72.5358N}
  {72, 5358}

\bibitem[\protect\citeauthoryear{{Oberst} et~al.,}{{Oberst}
  et~al.}{1998}]{1998M&PS...33...49O}
{Oberst} J.,  et~al., 1998, \mn@doi [Meteoritics and Planetary Science]
  {10.1111/j.1945-5100.1998.tb01606.x}, \href
  {http://adsabs.harvard.edu/abs/1998M%26PS...33...49O} {33}

\bibitem[\protect\citeauthoryear{{Palotai}, {Sankar}, {Free}, {Howell},
  {Botella}  \& {Batcheldor}}{{Palotai} et~al.}{2019}]{2019MNRAS.487.2307P}
{Palotai} C.,  {Sankar} R.,  {Free} D.~L.,  {Howell} J.~A.,  {Botella} E.,
  {Batcheldor} D.,  2019, \mn@doi [\mnras] {10.1093/mnras/stz1424}, \href
  {https://ui.adsabs.harvard.edu/abs/2019MNRAS.487.2307P} {487, 2307}

\bibitem[\protect\citeauthoryear{Parasuraman}{Parasuraman}{1986}]{parasuraman1986vigilance}
Parasuraman R.,  1986

\bibitem[\protect\citeauthoryear{{Racusin} et~al.,}{{Racusin}
  et~al.}{2008}]{2008Natur.455..183R}
{Racusin} J.~L.,  et~al., 2008, \mn@doi [\nat] {10.1038/nature07270}, \href
  {https://ui.adsabs.harvard.edu/abs/2008Natur.455..183R} {455, 183}

\bibitem[\protect\citeauthoryear{{Reddy}, {Dunn}, {Thomas}, {Moskovitz}  \&
  {Burbine}}{{Reddy} et~al.}{2015}]{2015aste.book...43R}
{Reddy} V.,  {Dunn} T.~L.,  {Thomas} C.~A.,  {Moskovitz} N.~A.,   {Burbine}
  T.~H.,  2015, {Mineralogy and Surface Composition of Asteroids}.
University of Arizona Press, pp 43--63,
  \mn@doi{10.2458/azu_uapress_9780816532131-ch003}

\bibitem[\protect\citeauthoryear{{Sansom}, {Bland}, {Paxman}  \&
  {Towner}}{{Sansom} et~al.}{2015}]{2015M&PS...50.1423S}
{Sansom} E.~K.,  {Bland} P.,  {Paxman} J.,   {Towner} M.,  2015, \mn@doi
  [Meteoritics and Planetary Science] {10.1111/maps.12478}, \href
  {http://adsabs.harvard.edu/abs/2015M%26PS...50.1423S} {50, 1423}

\bibitem[\protect\citeauthoryear{{Sansom}, {Rutten}  \& {Bland}}{{Sansom}
  et~al.}{2017}]{2017AJ....153...87S}
{Sansom} E.~K.,  {Rutten} M.~G.,   {Bland} P.~A.,  2017, \mn@doi [\aj]
  {10.3847/1538-3881/153/2/87}, \href
  {http://adsabs.harvard.edu/abs/2017AJ....153...87S} {153, 87}

\bibitem[\protect\citeauthoryear{{Sansom} et~al.,}{{Sansom}
  et~al.}{2019a}]{2019Icar..321..388S}
{Sansom} E.~K.,  et~al., 2019a, \mn@doi [\icarus]
  {10.1016/j.icarus.2018.09.026}, \href
  {https://ui.adsabs.harvard.edu/\#abs/2019Icar..321..388S} {321, 388}

\bibitem[\protect\citeauthoryear{{Sansom} et~al.,}{{Sansom}
  et~al.}{2019b}]{2019ApJ...885..115S}
{Sansom} E.~K.,  et~al., 2019b, \mn@doi [\apj] {10.3847/1538-4357/ab4516},
  \href {https://ui.adsabs.harvard.edu/abs/2019ApJ...885..115S} {885, 115}

\bibitem[\protect\citeauthoryear{See, Howe, Warm  \& Dember}{See
  et~al.}{1995}]{see1995meta}
See J.~E.,  Howe S.~R.,  Warm J.~S.,   Dember W.~N.,  1995, \mn@doi
  [Psychological Bulletin] {10.1037/0033-2909.117.2.230}, 117, 230

\bibitem[\protect\citeauthoryear{{Spurn{\'y}} \&
  {Borovi{\v{c}}ka}}{{Spurn{\'y}} \&
  {Borovi{\v{c}}ka}}{2013}]{spurny_2013_geminid}
{Spurn{\'y}} P.,  {Borovi{\v{c}}ka} J.,  2013, in {Masiero} J.,  ed.,
  Meteoroids Conference. \url
  {http://www.astro.amu.edu.pl/Meteoroids2013/main_content/data/abstracts.pdf}

\bibitem[\protect\citeauthoryear{{Spurn{\'y}}, {Haloda}, {Borovi{\v c}ka},
  {Shrben{\'y}}  \& {Halodov{\'a}}}{{Spurn{\'y}}
  et~al.}{2014}]{2014A&A...570A..39S}
{Spurn{\'y}} P.,  {Haloda} J.,  {Borovi{\v c}ka} J.,  {Shrben{\'y}} L.,
  {Halodov{\'a}} P.,  2014, \mn@doi [\aap] {10.1051/0004-6361/201424308}, \href
  {http://adsabs.harvard.edu/abs/2014A%26A...570A..39S} {570, A39}

\bibitem[\protect\citeauthoryear{Szegedy, Toshev  \& Erhan}{Szegedy
  et~al.}{2013}]{NIPS2013_5207}
Szegedy C.,  Toshev A.,   Erhan D.,  2013, in Burges C. J.~C.,  Bottou L.,
  Welling M.,  Ghahramani Z.,   Weinberger K.~Q.,  eds, , Advances in Neural
  Information Processing Systems 26.
Curran Associates, Inc., pp 2553--2561, \url
  {http://papers.nips.cc/paper/5207-deep-neural-networks-for-object-detection.pdf}

\bibitem[\protect\citeauthoryear{{Tonry} et~al.,}{{Tonry}
  et~al.}{2018}]{2018PASP..130f4505T}
{Tonry} J.~L.,  et~al., 2018, \mn@doi [\pasp] {10.1088/1538-3873/aabadf}, \href
  {https://ui.adsabs.harvard.edu/abs/2018PASP..130f4505T} {130, 064505}

\bibitem[\protect\citeauthoryear{Towner et~al.,}{Towner
  et~al.}{2020}]{2019arXiv190911883T}
Towner M.~C.,  et~al., 2020, \mn@doi [Publications of the Astronomical Society
  of Australia] {10.1017/pasa.2019.48}, 37, e008

\makeatother
\end{thebibliography}

\end{document}